\documentclass[10pt,journal,compsoc]{IEEEtran}
\ifCLASSOPTIONcompsoc
  \usepackage[nocompress]{cite}
\else
  \usepackage{cite}
\fi
\ifCLASSINFOpdf
\else
\fi
\usepackage{amsmath}

\usepackage{balance}

\usepackage{url}

\usepackage[utf8]{inputenc}
\usepackage{graphicx}
\usepackage{tikz}
\usetikzlibrary{shapes,arrows,quotes,positioning,calc,fit,backgrounds}

\usepackage{multirow}

\usepackage{subcaption}

\graphicspath{ {./images/} }

\definecolor{officegreen}{rgb}{0.0, 0.5, 0.0}

\usepackage{listings}
  \lstdefinelanguage{diff}{
	morecomment=[f][\color{diffstart}]{@},
	morecomment=[f][\color{diffincl}]{+},
	morecomment=[f][\color{diffrem}]{-},
        keepspaces=true,
  }

\makeatletter
\tikzset{
    database/.style={
        path picture={
            \draw (0, 1.5*\database@segmentheight) circle [x radius=\database@radius,y radius=\database@aspectratio*\database@radius];
            \draw (-\database@radius, 0.5*\database@segmentheight) arc [start angle=180,end angle=360,x radius=\database@radius, y radius=\database@aspectratio*\database@radius];
            \draw (-\database@radius,-0.5*\database@segmentheight) arc [start angle=180,end angle=360,x radius=\database@radius, y radius=\database@aspectratio*\database@radius];
            \draw (-\database@radius,1.5*\database@segmentheight) -- ++(0,-3*\database@segmentheight) arc [start angle=180,end angle=360,x radius=\database@radius, y radius=\database@aspectratio*\database@radius] -- ++(0,3*\database@segmentheight);
        },
        minimum width=2*\database@radius + \pgflinewidth,
        minimum height=3*\database@segmentheight + 2*\database@aspectratio*\database@radius + \pgflinewidth,
    },
    database segment height/.store in=\database@segmentheight,
    database radius/.store in=\database@radius,
    database aspect ratio/.store in=\database@aspectratio,
    database segment height=0.1cm,
    database radius=0.25cm,
    database aspect ratio=0.35,
}
\makeatother

\begin{document}
%
\title{Active Learning of Discriminative Subgraph Patterns for API Misuse Detection}
%
%
%
%

\author{Hong Jin Kang and
        David Lo
\IEEEcompsocitemizethanks{
\IEEEcompsocthanksitem H.J. Kang and D. Lo are with the School of Information Systems, Singapore Management University\protect\\
E-mail: hjkang.2018@phdcs.smu.edu.sg,davidlo@smu.edu.sg}
\thanks{}}

\newcommand{\tool}{ALP}

\IEEEtitleabstractindextext{%
\begin{abstract}
  A common cause of bugs and vulnerabilities are the violations of usage constraints associated with Application Programming Interfaces (APIs). 
API misuses are common in software projects,
and while there have been techniques proposed to detect such misuses, 
studies have shown that they fail to reliably detect misuses while reporting many false positives.
One limitation of prior work is the inability to reliably identify correct patterns of usage. 
Many approaches confuse a usage pattern's frequency for correctness.
Due to the variety of alternative usage patterns that may be uncommon but correct, 
anomaly detection-based techniques have limited success in identifying misuses.
We address these challenges and propose \tool{} (Actively Learned Patterns), 
reformulating API misuse detection as a classification problem.
After representing programs as graphs, \tool{} mines discriminative subgraphs. 
While still incorporating frequency information, through limited human supervision, we reduce the reliance on the assumption relating frequency and correctness.
The principles of active learning are incorporated to shift human attention away from the most frequent patterns. 
Instead, \tool{} samples informative and representative examples while minimizing labeling effort.
In our empirical evaluation, \tool{} substantially outperforms prior approaches on both MUBench, an API Misuse benchmark, and 
a new dataset that we constructed from real-world software projects.
\end{abstract}

}

\maketitle

\IEEEdisplaynontitleabstractindextext

%
\IEEEpeerreviewmaketitle

\IEEEraisesectionheading{\section{Introduction}\label{sec:introduction}}

In modern software projects, developers often depend on third-party libraries that provide reusable functionalities.
Third-party dependencies are accessed through their Application Programming Interfaces (APIs).
However, developers frequently use them incorrectly by violating usage constraints that these APIs may impose~\cite{egele2013empirical,nadi2016jumping,amann2016mubench}. 
Known as API misuses, the violation of these constraints is a frequent cause of bugs, resulting in software crashes and vulnerabilities.

There have been many techniques proposed to detect misuses of APIs~\cite{robillard2012automated}; both static and dynamic analysis have been employed.
However, recent studies have shown the ineffectiveness of existing techniques. 
Manually written specifications and API misuse detectors using static analysis have low precision and recall~\cite{legunsen2016good,amann2018systematic}.
From a survey of existing API misuses detectors by Amann et al.~\cite{amann2018systematic}, 
most existing detectors mine frequent patterns from existing code.
These detectors then look for code that deviates from these patterns, assuming that these deviations are API misuses. 
Along with recent studies~\cite{sven2019investigating,wen2019exposing,thummalapenta2011alattin}, Amann et al.~\cite{amann2018systematic} suggested that this naive assumption is the cause of the numerous false positives of existing detectors.
There may be uncommon patterns of usage that do not conform to the mined patterns,
but do not lead to bugs.
Furthermore, there are classes of APIs where their prevalent use are incorrect,
such as Java Cryptographic APIs~\cite{egele2013empirical,nadi2016jumping,kruger2019crysl,gao2019negative}. 
For these APIs, their correct usage may look like deviants of the most frequent patterns.
Due to these reasons, the frequency of a usage pattern is not a reliable signal of its correctness.

In this study, we propose that API misuse detection can be framed as a classification task. 
We propose our approach, \tool{} (Actively Learned Patterns), for detecting Java misuses.
While still using frequency information, \tool{} does not rely on it as the only signal of correctness, and incorporates 
some human supervision from an experienced user of each API to identify \textit{discriminative} subgraph features, 
i.e. subgraphs that occur more frequently in graphs of one label than the other.
These subgraph patterns act as indicators of either correctness or misuse.
Based on the discriminative subgraphs present in each usage example, they are classified with a machine learning classifier.

However, it is impossible to label every usage pattern of an API.
Furthermore, similar to findings from prior work~\cite{thummalapenta2011alattin}, 
we find that API usage is highly imbalanced, 
with most APIs exhibiting distributions where correct usage examples outnumber misuse examples.
Misuses are usually the minority class in this classification-based formulation of the API misuse detection problem.
While they are the minority, misuses may lead to bugs and vulnerabilities.
As a result, failing to detect them may have severe consequences.
A naive sampling of examples to label is likely to pick usage examples containing frequent patterns, 
while omitting less common but informative patterns that may be discriminative of the minority class.
Therefore, key to our approach is our careful use of usage examples from GitHub.
\tool{} uses active learning to identify informative examples for human annotation.
This enables \tool{} to sample a small but informative number of examples for labeling.

Next, to reduce misclassification, 
\tool{} withholds judgement on examples it cannot classify with confidence, 
such as uncommon usage patterns it has not seen, and leaves them for further human inspection.
Notice that this is in contrast with existing misuse detectors, 
which considers usage that deviates from previously seen examples as misuses.
Using a novelty detector, \tool{} withholds judgement on usage instances that it is uncertain of. 
During practical use, it is plausible for \tool{} to encounter usage instances of the API that do not resemble any training example. 
Given the prevalence of correct alternative usage patterns~\cite{thummalapenta2011alattin}, 
we suggest that the optimal behavior of an API misuse detection tool is to 
signal that it does not know how to classify such usage instances. 
By withholding judgement, \tool{} prevents wasted developer effort investigating false positives.

\tool{} has several features distinguishing it from previously proposed techniques:
\begin{itemize}
  \item \tool{}'s \textbf{graph classification} approach uses some supervision, and does not rely on the limiting assumption confusing frequently seen code as correct code.
  \item An \textbf{active learning}-inspired technique enables us to leverage the diversity of Big Code, 
        selecting informative and representative examples to label while minimizing human supervision.
      Thus, \tool{} learns subgraphs patterns that can characterize an API's usage, including uncommon usage patterns.
  \item Existing work assumes code deviating from mined patterns as misuses. We suggest that this is an invalid assumption and 
        we address deviations of known patterns through the use of classification with a reject option~\cite{chow1970optimum,herbei2006classification}.
      This is enabled by \textbf{novelty detection}, with which \tool{} refrains from providing judgement about usage instances it cannot classify  with certainty.
\end{itemize}

To evaluate \tool{}, we compared it against other static API misuse detectors for Java ~\cite{sven2019investigating,nguyen2009graph,monperrus2010detecting,wasylkowski2011mining,wasylkowski2007detecting} on an existing API misuse detection benchmark, MUBench~\cite{amann2016mubench}, 
as well as another dataset of 500 API usage examples constructed from 16 projects on GitHub.
Our evaluation results demonstrate the promise of the \tool{} approach.
On MUBench, 
\tool{} achieves a precision of 43.9\% compared to the state-of-the-art precision of 34.1\%, and a 
recall of 56.3\% compared to the state-of-the-art recall of 43.3\%.

To avoid bias from focusing on the misuses collated in MUBench, which were previously identified by existing misuse detectors, 
we construct a new dataset, \textit{AU500}. 
Labeled independently by multiple human annotators, the AU500 has 500 usage instances randomly sampled from all API usage instances 
in 16 projects studied in prior work~\cite{wen2019exposing}.
On the AU500, \tool{} achieves 44.7\% precision and 54.8\% recall
outperforming
the state-of-the-art tool, MUDetectXP~\cite{sven2019investigating}, which achieves 27.6\% precision and 29.6\% recall.
While we do not solve the challenges of automated API misuse detection in their entirety, 
the substantial improvements shows that \tool{} is a step forward. 
The contributions of this study are:
\begin{itemize}
  \item \textbf{Problem Formulation:} We formulate API misuse detection as a graph classification task, identifying discriminative subgraphs as features.  
        As far as we know, this is the first work to propose the combination of graph classification and active learning on an API-related problem.
  \item \textbf{Approach}: We propose \tool{}, a novel approach utilizing active learning-based sampling and a reject option for classification, 
  addressing challenges faced by existing approaches.
  \tool{} and the AU500 are available on its artifact website\footnote{\url{https://github.com/ALP-active-miner/ALP}}.
  \item \textbf{Evaluation}: Our evaluation results show the promise of \tool{} in API misuse detection, outperforming prior work on MUBench 
  and a new dataset we construct, the AU500. 
  While MUBench has labels only for the 208 misuse locations detected by prior techniques, 
  the AU500 dataset has 500 labeled usage sites of both correct usage and misuses.

\end{itemize}

\section{Background}
\label{sec:background}
We present some background information of \tool{}. 
Through examples, we motivate the features of \tool{} through the challenges of detecting misuses that are tackled by \tool{}. 
Afterwards, we describe the API Usage Graph~\cite{sven2019investigating}, a graph representation of source code 
shown to be promising for the API misuse detection problem. 
We build \tool{} on top of the API Usage Graph.

\subsection{Motivating Examples}

In this section, we show simplified examples from GitHub projects that show the challenges we attempt to address.
Existing misuse detectors usually conflate frequent usage patterns and correct usage patterns, then look for 
API usage instances that deviate from these patterns.
The two primary challenges that we tackle in this work are that there is no known way to reliably and automatically evaluate 
\textbf{the correctness of a mined pattern (Challenge 1)} 
and \textbf{the correctness of a usage example that deviates from known usage patterns (Challenge 2)}. 

\begin{figure}[h]
	\centering
\begin{lstlisting}[language=java,numbers=none,basicstyle=\ttfamily,frame = single]
Cipher cipher = Cipher.getInstance("DES");
cipher.init(1, secretKey);
byte[] textBytes = text.getBytes(charset);
byte[] bytes = cipher.doFinal(textBytes);
\end{lstlisting}
    \caption{A usage pattern involving Cipher. This uses the "DES" algorithm, known to be insecure~\cite{egele2013empirical}.}
    \label{fig:cipher_alternative_patterns}
\end{figure}

The distribution of API usages is highly skewed, and most frequency-based pattern mining approaches assume that the most common patterns are more likely to be correct.
However, this is not the case for some APIs, such as cryptographic APIs.
An example of an incorrect API usage is shown in Figure \ref{fig:cipher_alternative_patterns}.
Hence, the assumption that a highly frequent usage pattern is a correct usage is not a valid assumption,
and we address this by including some human supervision.
Still, even with a user of the APIs labeling a sample of usage examples, it is challenging to train an effective classifier.  
Due to the imbalanced distribution of API usage patterns,
if one tries to learn usage patterns through a random sampling of examples from GitHub,
then it is possible that the sample does not contain a single example using an uncommon pattern.
In the case of \textbf{Cipher}, one may only encounter examples similar to the above, 
and miss out the uncommon but correct usage examples.

The second challenge is related to the high rate of false positives.
To detect misuses, existing tools look for deviations from mined patterns.
Simply reporting any deviation as a misuse will produce numerous false positives, 
and 
the tools mitigate this problem through various heuristics.

For an example API, we look at \texttt{java.util.Map}. 
Included in Java's standard library, this data structure is ubiquitous in Java projects. 
Yet, when we inspect the evaluation results of existing detectors on API misuse detection benchmark, MUBench~\cite{amann2016mubench,amann2018systematic}, 
we find that none of the existing misuse detectors are able to detect misuses of \texttt{Map}.
We hypothesize that this is caused by the variety of usage contexts it appears in.
Based on an empirical analysis by sampling API usage instances from GitHub, the idiomatic usage pattern 
is a null-check performed on the return value of the \textit{get(key)} method call,
but there are less common patterns that may not be idiomatic, but are safe and correct.

\begin{figure}[h]
	\centering
	\scriptsize{
\begin{lstlisting}[language=java,numbers=none,basicstyle=\ttfamily,frame = single]
  // 1. A pattern that is common and correct
  map1 = new HashMap<>();
  Integer val1 = map1.get("data");
  if (val1 != null) {
    ...
  }
  // 2. A pattern that is uncommon but correct.
  // If reported as a misuse, it would be a false positive 
  map2 = new ConcurrentHashMap<>();
  for (String key : map2.keySet()) {
    int val2 = map2.get(key);
    ...
  }
  
  // 3. A misuse of Map. 
  // If not reported as a misuse,
  // it would be a false negative.
  // codeMap.get() may return null
  ((Integer)this.codeMap.get(
    cause.getClass())).intValue()
\end{lstlisting}
    \caption{Three usages of Map. The first is common and correct, while the second is uncommon but correct. The third is a simplification of a misuse from MUBench.}
    \label{fig:map_alternative_patterns}
	}
\end{figure}

In the second example shown in Figure \ref{fig:map_alternative_patterns}, 
we see an uncommon usage of a \texttt{ConcurrentHashMap}.
While it implements the Map interface, \texttt{ConcurrentHashMap} does not allow for null values in it.
Hence, invoking \textit{get} on keys of the map (by iterating over \textit{keySet}) never returns null.
However, based on the frequent pattern of a null-check, this usage instance may be considered a misuse as it deviates from the pattern.
Reporting it results in a false positive.
Many tools use a ranking-based approach by computing various metrics,
such as \textit{rareness}~\cite{nguyen2009graph}, 
to rank their output with the aim of ranking false positives below true positives.

The third usage is a misuse, in which a method is invoked on the return value of \textit{get} without checking for null, 
but is not detected by existing techniques.
In particular, while it is a deviation from the null-check pattern, 
MUDetectXP does not report such usage instances as it heuristically omits fields from its findings.
This restriction was required to prevent numerous false positives related to the usage of fields.
While this  heuristic succeeds in reducing the number of false positives, it may cause MUDetectXP to miss misuses.

The poor empirical performance~\cite{amann2018systematic} of these tools suggest the need for a different approach to reduce false positives.
We suggest that, despite some success in using heuristics to prevent the reporting of false positives, 
the fundamental problem of distinguishing real usage constraints from spurious usage patterns remains unsolved.

We also address a third challenge, which is that \textbf{API usage patterns can be complex (Challenge 3)}.
Some studies model method calls and their order of invocation, data-flow, and control-flow~\cite{wasylkowski2007detecting,wasylkowski2011mining}, 
other studies has suggested the need for modelling multiple objects~\cite{pradel2012statically},  
static methods and constants~\cite{zhong2017empirical},
self-usages~\cite{sven2019investigating} (the usage of an API within its own implementation), 
inheritance~\cite{zhong2017empirical},
argument values~\cite{egele2013empirical,zhong2020para}, 
and synchronization~\cite{lin2016lockpeeker,sven2019investigating}.
Consequently, the representation of a usage pattern has to be sufficiently rich to capture this complexity.

\begin{figure}[h]
  \centering
  \scriptsize{
\begin{lstlisting}[language=java,basicstyle=\ttfamily,numbers=left,frame = single,framexleftmargin=2.1em, xleftmargin=2.0em]
class CopyOnWriteMap<K,V> implements ConcurrentMap<K,V> { 
  ...
  public V putIfAbsent(K k, V v) {
    synchronized(this) {
      if (!containsKey(k)) 
        return put(k, v);
      else 
        return get(k);
    }
  }
  ...
  public V get(K k) {
    this.internalMap.get(k);
  }
}

\end{lstlisting}
    \caption{Example usage of Map in a CopyOnWriteMap, which transitively implements the Map interface}
    \label{fig:aug_code}
  }
\end{figure}

In Figure \ref{fig:aug_code}, we show an example of the complex relationships between program elements related to usage of an API.
We see uses of \texttt{Map} that have relationships with program elements 
beyond control-flow. 
The class, CopyOnWriteMap, is a sub-type of \texttt{Map} through the \texttt{ConcurrentMap} interface. 
The \textit{get(key)} on line 8 is a self-usage, invoking \textit{get(key)} which is overridden by the class (line 12). 
The object, itself a \texttt{Map}, is used to synchronize access to the other method calls on line 4.

\subsection{API Usage Graph}
\label{sec:aug}

\begin{figure}[]

  \includegraphics[width=\columnwidth]{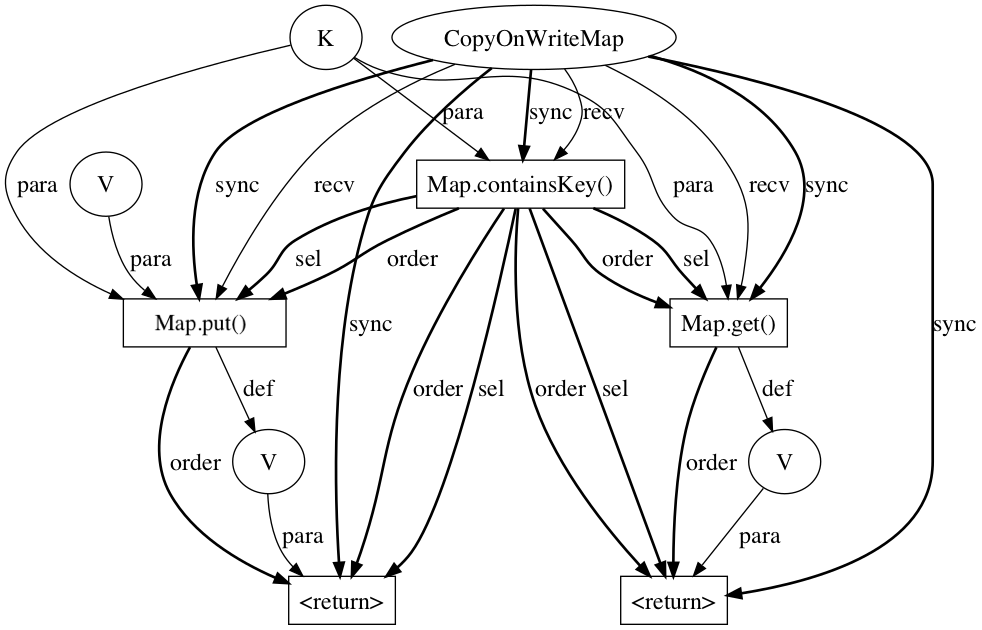}
\centering
\caption{Example of an API Usage Graph}
  \label{fig:aug}
\end{figure}

As prior work~\cite{amann2018systematic,nguyen2009graph,sven2019investigating,zhong2017empirical} have shown the promise of using a graph representation of API usage for detecting API misuses,
our work represents programs as graphs. 
Amann et al.~\cite{sven2019investigating} proposed the  API Usage Graph.
An example is shown in Figure \ref{fig:aug}, constructed based on the \textit{putIfAbsent} method in Figure \ref{fig:aug_code}. 
The API Usage Graph is a directed multi-graph with labeled nodes and edges.
Nodes represent objects, values, method invocations, constructor calls, field accesses, and conditional checks.
Edges represent data and control flow between the program elements of the nodes.
Edges have the following labels: 
\begin{itemize}
\item \texttt{recv} (linking method calls that are invoked on a object), 
\item \texttt{param} (linking variables/literals used as arguments to the methods called. While labeled “param”, a more accurate label may be “arg” as this links variables/literals used as arguments to method call.), 
\item \texttt{def} (linking actions creating/returning a data), 
\item \texttt{order} (linking actions in order of execution), 
\item \texttt{sel} (for control-flow; representing a control-relationship), 
\item \texttt{sync} (linking actions to objects they are synchronized on), 
\item \texttt{throw} and \texttt{handle} (for exceptional flow).
\end{itemize}

Amann et al.~\cite{sven2019investigating} mined subgraphs of the API Usage Graph. 
Alternatives to using the API Usage Graph are to use patterns mined from other representations of a program. 
For example, Grouminer~\cite{nguyen2009graph} considers control and dataflow dependencies, 
while DMMC~\cite{monperrus2010detecting} encodes the set of methods called on an object.
Still, the expressivity of the API Usage Graph allows it to distinguish more types of misuses from correct usage. 
While previously proposed detectors have modelled some of these relationships, 
they were not considered in tandem. 
On the other hand, these relationships are combined in the API Usage Graph~\cite{sven2019investigating}.
Therefore, patterns mined  in these prior studies~\cite{nguyen2009graph,monperrus2010detecting} can also be represented as a subgraph pattern in the API Usage Graphs
Frequent subgraphs mined from API usage graphs was shown to be able to capture usage constraints of APIs in previous work ~\cite{sven2019investigating}.
Therefore, we express programs as API Usage Graphs in our work and mine subgraphs from them.

\subsubsection{Discriminative Subgraph Mining}

Many existing approaches focus on mining frequent usage patterns of source code using an API.
Frequent subgraphs are discovered by identifying subgraphs occuring more than a user-specified number of times in a collection of graphs.
Most of these frequent subgraphs will not be discriminative of API misuses,
in other words, they may appear equally frequently in both correct and incorrect API usages.
Moreover, there is usually a large number of subgraphs with frequency greater than the user-specified number of times.
The large volume of frequent subgraphs may make further processing of the subgraphs unscalable.

One solution to the limitations of frequent pattern mining  is to mine discriminative subgraphs.
Used for graph classification tasks where graphs have different labels, 
these subgraphs are both frequent and have discriminative power to distinguish between graphs of different labels.
In our study, we mine discriminative subgraphs to distinguish API usage graphs that are misuses from API usage graphs that are correct.
In other words, we look only for frequent subgraphs that are indicative of either correct or incorrect API usage.
While all discriminative subgraphs are frequent, not all frequent subgraphs are discriminative.

\begin{figure}[]

  \includegraphics[width=\columnwidth]{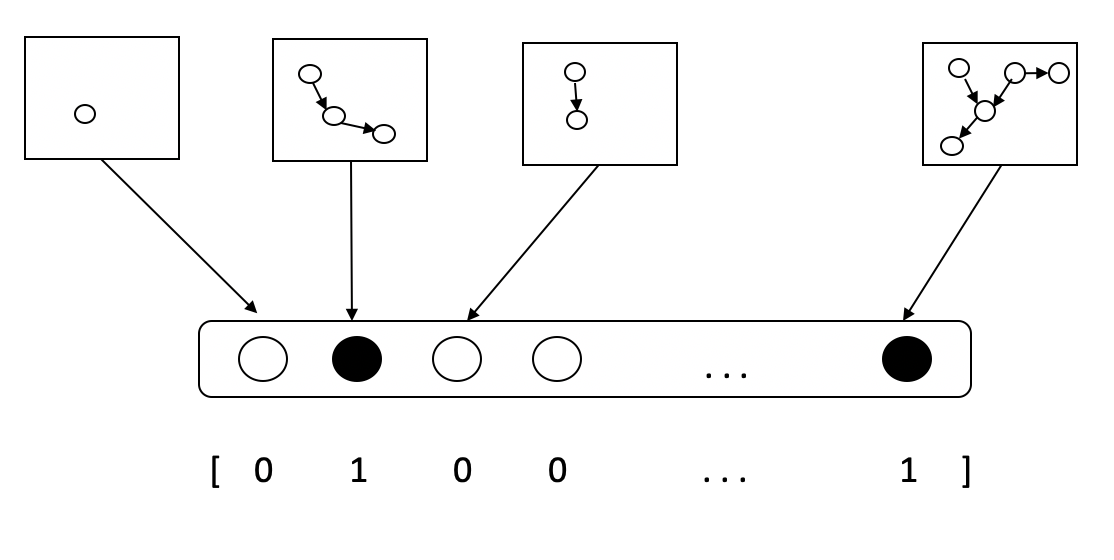}
\centering
\caption{Vector representation of a usage example. The \textit{i}'th element in the vector is 1 if the \textit{i}'th discriminative subgraph is present in the usage example. Given that the second element represents a check for \texttt{containsKey}, and the program in Figure \ref{fig:aug_code} contains a checks for \texttt{containsKey}, the second element is 1. }
  \label{fig:vector}
\end{figure}

Graphs can be represented by the subgraphs that they contain.
To use discriminative subgraphs to perform classification,
each usage location is expressed in terms of the presence and absence of these subgraphs.
A vector of the same length of the number of discriminative subgraphs, containing 0s and 1s representing absence and presence respectively, is passed into a machine learning classifier.
For example, the source code given in Figure \ref{fig:aug_code} could be represented as a vector in Figure \ref{fig:vector}.
A single usage example may match multiple subgraph features (e.g. both the second and last subgraph features in Figure \ref{fig:vector} are matched).
The other discriminative subgraphs that are not present in the API usage location have a value of 0 in the vector.
Representing the usage in this form allows the use of machine learning classifiers which take vectors as input.

The task is of a probabilistic nature; 
while it is typically the case that \textit{containsKey} followed by \textit{get} is a correct usage,
this usage pattern does not guarantee that the usage is correct. 
Some implementations of \texttt{java.util.Map} may allow a \texttt{null} value.
As such, an approach that directly matches programs to a singular pattern cannot capture this uncertainty
while the machine learning classifier-based approach of \tool{} reflects the nature of this task.

\subsection{Active Learning}

Active learning is a subfield of machine learning that aims to achieve better effectiveness while requiring fewer labelled examples~\cite{dagan1995committee,lewis1994sequential,lewis1994heterogeneous,cohn1996active,fujii1998selective}.
Many machine learning techniques require many labeled examples, 
which are examples where their true class has been indicated by a human annotator, 
to train a model. 
Active learning techniques can be useful in situations where labels of examples are hard to obtain.
Rather than selecting random examples to be labelled, 
active learning aims to select informative and representative examples, 
and has been shown to be effective in minimizing the number of labels required to learn an effective classifier.
Approaches using active learning ask \textit{queries} that are answered by an \textit{oracle} (usually human annotators).
These queries are unlabelled examples;
their true classes are not known yet as they have not been labeled by the human annotator. 
In the context of graph classification, an active learning technique poses \textit{query graphs}, unlabeled graphs, 
to the human annotator.

\begin{figure}[]

  \includegraphics[width=\columnwidth]{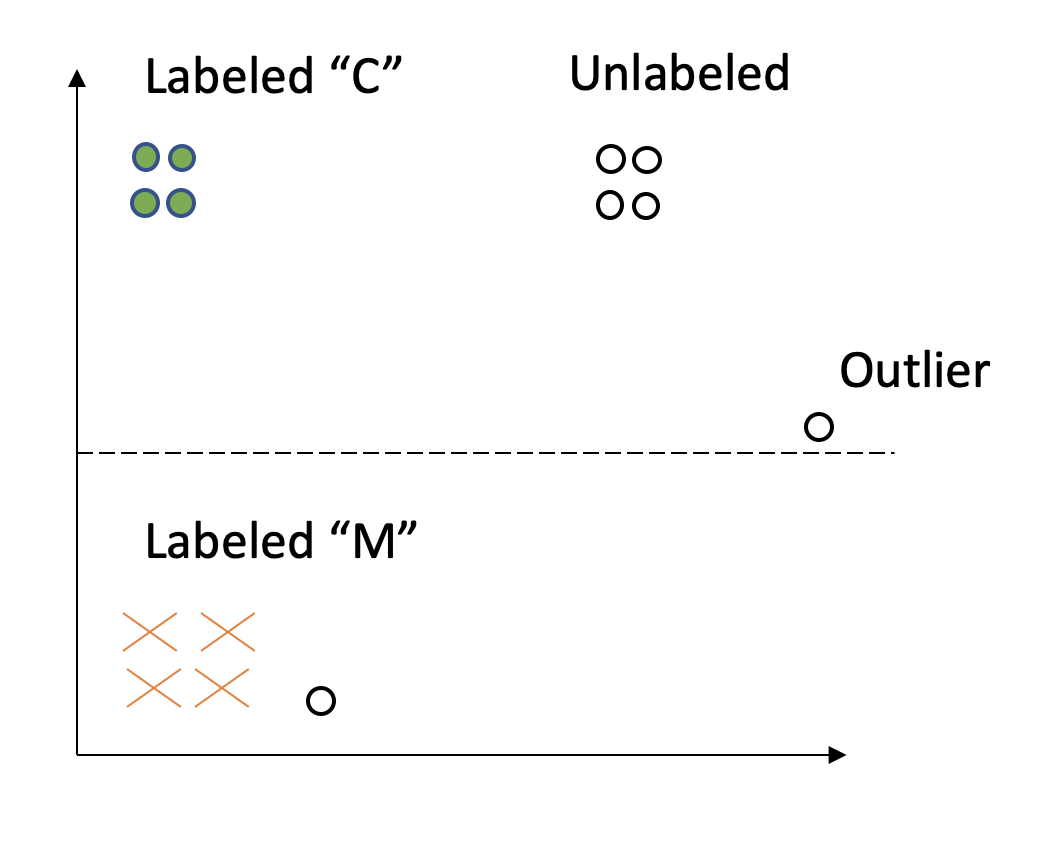}
\centering
\caption{Data points projected onto the input feature space. Each point represents one example. The dashed line represented the decision boundary of the model, which learned to distinguish between examples of two different labels. The figure indicates two groups of examples that have been labeled. The unfilled circles indicate examples that have not been labeled. An active learner may select an example from Unlabeled for labeling.  }
  \label{fig:clusters}
\end{figure}

Many frameworks for querying unlabelled examples have been proposed~\cite{dagan1995committee,lewis1994sequential,lewis1994heterogeneous,cohn1996active,fujii1998selective}. 
Many of these strategies direct the human annotator’s attention at \textit{informative} examples, 
such as examples where the model produces the most uncertain predictions~\cite{dagan1995committee,lewis1994sequential,lewis1994heterogeneous},
examples that would cause the model to change the most~\cite{settles2008analysis},
or examples that will reduce the variance of a model~\cite{cohn1996active}.
In Figure \ref{fig:clusters}, points with similar features as the already labeled examples (and are close to them in the input space) 
are less likely to be selected, 
since they likely share the same label as the labeled examples.
An example from the \textbf{Unlabeled} region is likely to be selected as it is far from the already labeled examples.
Another direction of research measures the \textit{representativeness} of unlabelled examples,
addressing limitations of prior techniques that tend to select outliers for labeling (e.g. selecting the outlier in Figure \ref{fig:clusters} may not give us information about the other points in the  \textbf{Unlabeled} cluster).
These techniques, e.g. \cite{settles2008analysis,fujii1998selective}, favour examples that are in dense regions of the input space,
or are most similar to other unlabelled examples (e.g. examples from \textbf{Unlabeled} in Figure \ref{fig:clusters})~.
In short, 
active learning tries to minimize the total number of labels required by selecting query examples that are informative and representative.
An informative example is one that should be dissimilar to examples that are already labelled, 
while a representative example is one that is similar to other unlabelled examples.

In \tool{}, we leverage active learning to iteratively identify examples from GitHub to be labelled by the human annotator.
We describe in Section \ref{sec:selection} how we determine unlabelled usage examples that are dissimilar 
from already labelled examples, while ensuring that they are similar to other unlabelled usage examples.

To motivate this, 
we use Figure \ref{fig:cipher_alternative_patterns} as an example.
If \tool{} has already obtained enough examples to mine a feature indicating that \textit{getInstance("DES")} is an incorrect use,
\tool{} will focus more on examples that do not contain \textit{getInstance("DES")} (hence, dissimilar to existing labeled examples). 
However, of the different usage examples, it will be helpful to focus on examples with usage patterns that 
do not appear by coincidence and are not outliers (in other words, examples which are similar to other unlabelled examples).
These heuristics would help \tool{} to locate examples that use other types of algorithm (e.g. \textit{"AES"}).

\subsection{Learning with Rejection and Novelty Detection}

Researchers have proposed a framework for classification with a reject option~\cite{chow1970optimum,herbei2006classification,cortes2016learning}. 
This framework consists of two components: a traditional classifier and a rejection function.
The goal is a machine learning model that knows what it does not know.
It is suggested that this framework is useful for scenarios where incorrect predictions can be costly,
such as medical diagnosis~\cite{herbei2006classification}.
Typically, approaches that incorporate learning with rejection estimate the confidence of a prediction.
If less confident about a prediction, the model withholds its prediction.

Rejecting a classification and task of novelty detection are closely related~\cite{tax2008growing}; 
a classification should be rejected if the object under classification is an outlier~\cite{tax2008growing}, 
and 
novelty detection is the task of identifying that some test example does not resemble the examples used during training. 
While modern machine learning techniques learn from large amounts of data, it is still possible that a large number of examples 
that exist in the real world are not reflected within the dataset. 
Solutions to this problem learn a model of ``normal'' examples seen during training, 
and use the model to detect ``abnormal'' examples during testing. 
One-class classification can be used to approach this problem~\cite{gardner2006one,perera2019ocgan}, in which a model learns to differentiate the ``normal'' class seen during training 
from all other examples~\cite{khan2014one}.

In \tool{}, we use a classifier with a reject option, in which 
we use an off-the-shelf novelty detector (later described in Section \ref{sec:classifying}) as a measure of the confidence of a prediction.
If the novelty detector determines that the example is novel, then it rejects the classification and does not make a prediction.
The novelty detector compares a test instance to its neighbours to determine how isolated the instance is;
the more isolated and further it is from its neighborhood, the more novel it is. 

\begin{figure}[]

  \includegraphics[width=0.9\columnwidth]{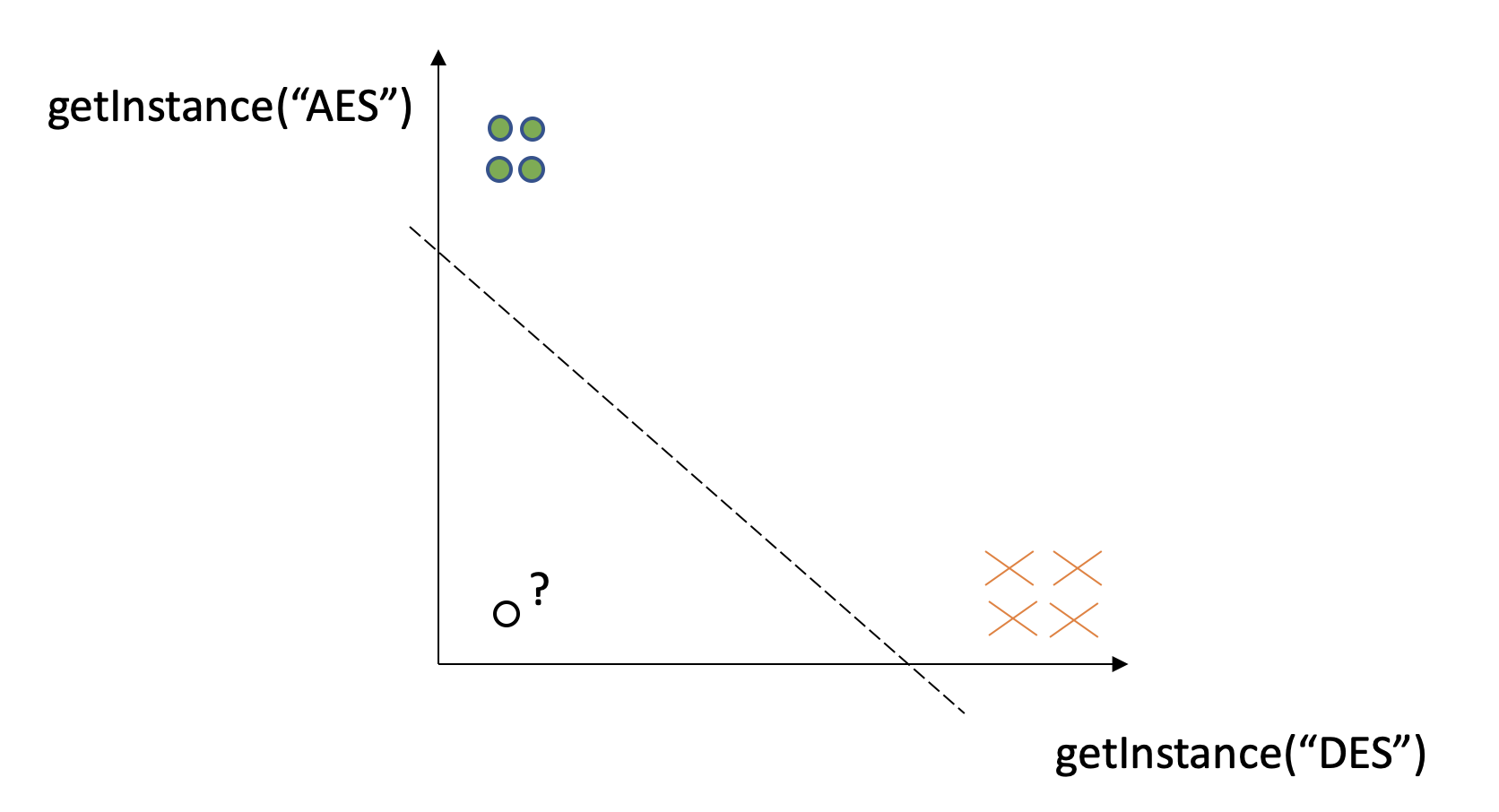}
\centering
\caption{Examples projected onto the input feature space. Given that two features related to ``DES'' and ``AES'' have been identified, a novelty detector detects if a test instance (shown with a question mark, ``?'') uses an algorithm it has not seen (e.g. ``RSA''). This allows \tool{} to make the right choice to reject the classification as it has no way to make the right prediction. }
  \label{fig:clusters}
\end{figure}

In the example of \texttt{Cipher}, 
if \tool{} has only seen examples that allowed it to mine some features for a small set of algorithms 
(e.g. \texttt{getInstance("DES")}  and \texttt{getInstance("AES")}), 
it will not be able to correctly judge the correctness of a previously unseen algorithm.
When faced with a test instance that use a new algorithm, e.g. \texttt{getInstance("RSA")}, 
as shown in Figure \ref{fig:clusters} as the question mark (``?''), 
the novelty detector allows \tool{} to detect that it is a an out-of-distribution instance as
it does not resemble any of the training examples, all of which have one of the two subgraph features.

\section{Approach}

\begin{figure*}[htb]
  \begin{center}

  \includegraphics[width=1.9\columnwidth]{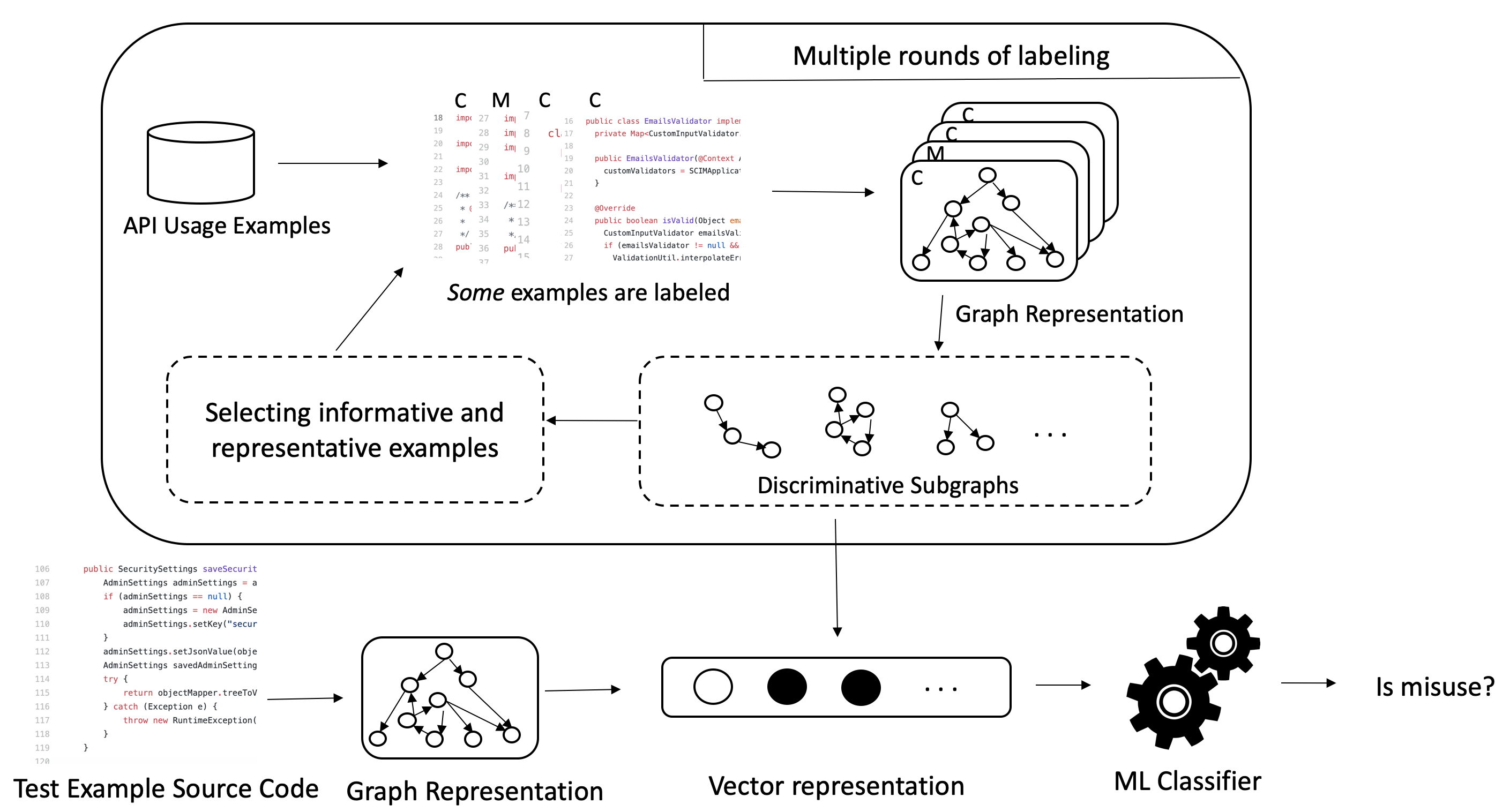}
  \end{center}
  \caption{High-level overview of \tool{}. \tool{}'s objective is to classify API usages. A user labels a small sample of usage examples, categorizing them as misuses or correct uses.
  Discriminative subgraphs, which are indicative of either correct uses or misuses, are identified from these labeled examples.
  ALP may identify more samples of informative and representative code examples for labeling.
  Based on the discriminative subgraphs, a given API usage can be represented as a vector, and then input to a Machine Learning classifier to determine if it is a misuse.
  }
  \label{fig:approach}
\end{figure*}

\subsection{High-level Overview}

In \tool{}, the API misuse detection task is reformulated as a graph classification task. 
In this formulation of the problem, usage sites of an API (methods using an API) can be classified as either a misuse or a correct usage.
Summarized in Figure \ref{fig:approach},
\tool{} relies on a human annotator to label a small set of selected examples as correct and misuse examples.
Next, it performs discriminative subgraph mining 
while picking more examples for the human annotator to label based on principles of Active Learning. 
Finally, a machine learning classifier is trained together with a novelty detector. 

We represent each method with a usage of the API as a graph, $G$. 
\tool{} extends the API Usage Graph to represent source code (Section \ref{sec:eaug}).
Examples of API usage are mined from GitHub (Section \ref{sec:github}) and these examples are transformed 
into graphs.
In order to train and use a machine learning classifier, each graph is represented as a feature vector, $v$. 
To this end, we identify and use discriminative subgraphs features (Section \ref{sec:subgraph_mining}). 

The difficulty of addressing \textbf{Challenge 1} using a supervised approach is the need to 
minimize human effort while labeling examples that are less common but informative.
It is not possible for a human to label every possible usage pattern.
\tool{} directs human attention to examples that differ from already labeled examples to maximise the diversity of labeled examples,
addressing \textbf{Challenges 1 and 2}.
This is enabled by an active-learning inspired approach (Section \ref{sec:selection}), in which  
\tool{} iteratively queries the human annotator for the labels of a small but informative batch of examples.

Finally, we perform classification using a machine learning classifier (Section \ref{sec:classifying}).
We utilize reject option in classification~\cite{chow1970optimum,herbei2006classification} for handling usage instances with uncertain labels. 
This helps addresses \textbf{Challenge 2} during testing time,  
detecting if the usage instance is too abnormal from training usage examples.
When faced with an abnormal example, \tool{} does not consider it a misuse, but defers judgement.
The classification module outputs one of three labels, Rejection (signalling uncertainty), Misuse, or Correct.

\tool{} represents programs using the API Usage Graphs (see Section \ref{sec:aug}).
This enables \tool{} to mine subgraph features that are complex, 
consisting of multiple types of information 
(e.g. both constraints on parameter values and control-flow can be included within a single pattern).
To tackle \textbf{Challenge 3}, we address several limitations of the API Usage Graph by extending it to include richer information about the program. 
In this work, we refer to our extension as the EAUG (Extended API Usage Graph).

To sum up, \tool{} consists of the following components:
\begin{itemize}
  \item \textbf{Extensions to the API Usage Graph (Extended API Usage Graph)}. This allows for the subgraph mining process to mine features that are key to avoid false positives.
  \item \textbf{A wrapper over GitHub's search API}. This is used to obtain usage examples.
  \item \textbf{A loop over the human annotating the examples, the discriminative subgraph miner, and the selector of examples to label.} This is the only component requiring (potentially multiple iterations of) human input.
  \item \textbf{The graph classifier.} Based on the features mined by the discriminative subgraph miner, the classifier is tuned over the training dataset. It outputs one of three labels (\textbf{C}orrect, \textbf{M}isuse, and \textbf{U}nknown) for an unlabelled instance during testing time. 
\end{itemize}

\subsection{Workflow of \tool{}}

\tool{} is intended for use by an experienced user of a given API. 
We assume that these users are knowledgeable about the use of the API, and can accurately label the correctness of their usages with relatively low effort.
ALP enables a user to identify potential problems without writing a specification by hand.
Some studies have shown that manually written specifications may introduce many false positives as they fail to account for all usage patterns used by API clients~\cite{legunsen2016good}.
Other studies also suggest that developers face many challenges writing specifications~\cite{chalin2006practitioners,schiller2014case}.

In this work, the first author studied the APIs carefully before labeling the data.
We envision that our work may be useful for an API developer to locate all projects and locations where the API may be used incorrectly, 
such as if a vulnerability related to the API has been reported, 
or if the API developer plans to introduce a breaking change  
that may affect clients that have used the API in a way not expected by the API developer~\cite{hyrumslaw} (e.g. due to API workarounds~\cite{lamothe2020apis}).
There is often a knowledge gap between API developers and API users~\cite{lamothe2020apis,robbes2011study}, 
and it is possible that users may use an API in ways that violate undocumented usage constraints (i.e., misuses) .
For such cases, a modification to the API by the developer may break the client projects~\cite{moller2020detecting,foo2018efficient}.
Considering the latter usage scenario (developer introduces breaking changes), the following illustrates the benefit of ALP:

\textbf{Without ALP}, the API developer will have to search for all usages of the API 
and manually inspect them to determine if it is a misuse. 
The API developer may try to filter the usages, 
but ultimately will find it difficult to perform a search that checks if a given constraint holds and to enumerate over all the valid usage scenarios. 
The developer may also miss out uncommon usage patterns. This leads to loss of time and increases the cost of maintenance when the API developer wishes to introduce breaking changes. 

\textbf{With ALP}, the API developer only has to label a small number of usage examples, 
and ALP can classify the locations that are likely to be misuses, 
acting as a filter for usages that the developer has to inspect. 
Moreover, once ALP has been trained, it can be used again without any cost. 
The developer only needs to inspect the few locations that ALP reports misuses in, 
and can quickly reach out to the developers of the client projects. 
As a result, the developer can save time and effort while detecting potential problems.

The workflow of using \tool{} is given as follows:
\begin{itemize}
  \item First, the type of the API is given as input, 
  then \tool{} randomly selects a small number of examples for the user to label.
  \item Next, the first batch of inputs is input to \tool{}, which may query the user for the labels to another batch of examples. This step is repeated until the stopping criteria (e.g. a maximum of 5\% of the dataset are labelled, or if \tool{} has found enough discriminative subgraphs for 95\% of the dataset) is met. After this step, no further data is labeled.
  \item \tool{} uses the labeled examples as input to construct a machine learning model of the API. Both the graph classifier and novelty detector are trained in this step. 
  \item After \tool{} has a trained model of the API, it can accept test instances as input.
        Given a file with source code using an API, \tool{} converts each method using the API into a vector.
        For each instance, the model produces one of three labels, \textbf{C}orrect, \textbf{M}isuses, or \textbf{U}nknown.
\end{itemize}

Within this study, we limit ourselves to look for misuses in MUBench and AU500.
To evaluate the generalizibility of the approach, 
we do not label usage examples from projects that are present in MUBench and the AU500.
The usages in MUBench and AU500 are used only as test instances.

\subsection{Extended API Usage Graph}
\label{sec:eaug}

To mine discriminative subgraphs, we represent an API usage as a graph, 
and we propose extensions to the API Usage Graph (AUG).
While the AUG can represent many important aspects of an API usage, 
we find that it considers only information of program elements directly related to the execution of the program.
We hypothesize that other elements in the source code, which may not directly influence a program's execution,
may serve to inform developers of their purpose and 
can act as indicators of the different contexts of an API usage.
Extending the AUG with these indicators may allow subgraphs that are more discriminative to be mined. 
In this subsection, we motivate the choices we make for extending the AUG using concrete examples from GitHub.

\begin{figure}[h]
  \centering
  \scriptsize{
\begin{lstlisting}[language=java,basicstyle=\ttfamily,numbers=left,frame = single,framexleftmargin=2.1em, xleftmargin=2.0em]
void sendNck(String protocol, PrintWriter printWriter, 
             String result) {
  printWriter.write(protocol + TOKEN_DIVISION +
                    result + "\r\n");
  printWriter.flush();
}

void quit(String nickName, PrintWriter printWriter,
          BufferedReader bufferReader, Socket socket) {
  ... // removed irrelevant code
  printWriter.close();
  ... // removed irrelevant code
}

\end{lstlisting}
    \caption{Simplified example usage of \texttt{PrintWriter}. The \texttt{PrintWriter}'s use is spread across multiple methods.}
    \label{fig:printwriter}
  }
\end{figure}
\vspace{0.2cm}\noindent{\bf Parameters and fields.} The AUG does not distinguish between variables that are instantiated locally in the scope of the current method, the fields of the class, and the parameters of the current method.
We suggest that having the ability to distinguish between them will help to further distinguish between different usage contexts, 
including self-usages~\cite{amann2018systematic} where the implementation of an API calls itself. 
In practice, developers may use fields and parameters differently from local variables.
Fields have a wider range of usage compared to local variables; 
the usage of fields may extend beyond a single method and may hold the result of partial computation, 
may be used for synchronization, or for caching.
Empirically, we found that the use of some APIs are not self-contained within a single call site, 
e.g. many usages of \texttt{PrintWriter} use it as a field in an enclosing class.
While the correct use of a \texttt{PrintWriter} requires \textit{close} to be invoked at the end of its lifecycle,
its correct use may be spread across multiple methods of the enclosing class.
Hence, a misuse only occurs if a usage constraint of the enclosing class is, itself, violated.
Another frequently observed pattern (e.g. in Figure \ref{fig:printwriter}) 
is that a \texttt{PrintWriter} is passed as an argument to the method,
and it would not be a misuse if the \texttt{PrintWriter} wasn't closed as 
the client of the API that passed in the \texttt{PrintWriter} should close it from the outside of the API.
On the other hand, had the \texttt{PrintWriter} been instantiated as a local variable, 
then failing to invoke \textit{close} on it will cause a resource leak.
Note that while the AUG has an edge type (described above as the \texttt{param} edge) to indicate that a particular variable is passed as an argument in a method call,
the AUG does not indicate the parameters of the method represented by the AUG. 

In the EAUG, 
we indicate fields and parameters differently from local variables, indicating this piece of information in the data node.
Similar to fields, method parameters may have implicit usage constraints on them and may follow specific rules and patterns~\cite{zhong2020para,zhang2012automatic}.

\begin{figure}[]

  \includegraphics[width=\columnwidth]{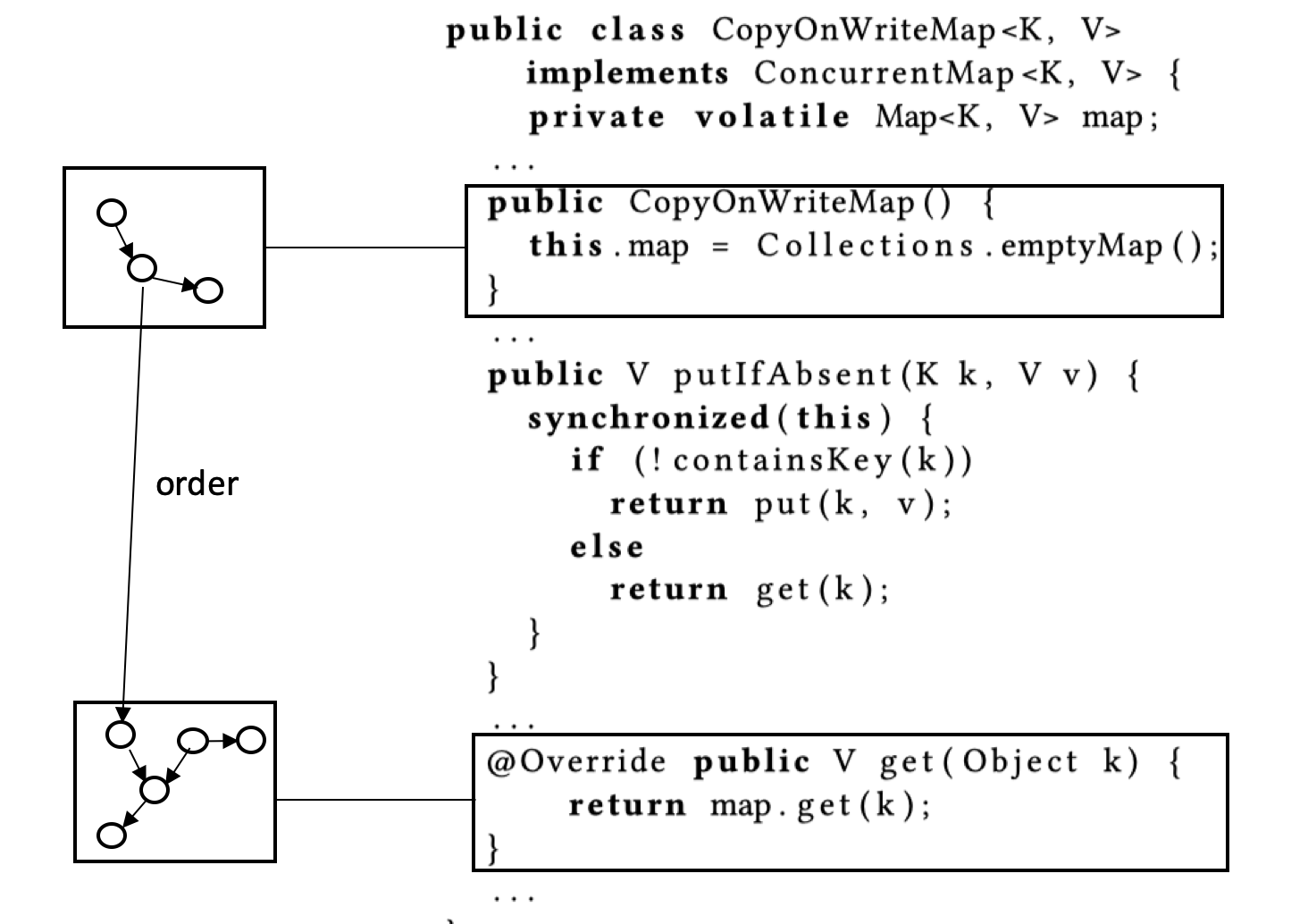}
\centering
\caption{When a field is used in a method, the initialization of the field is linked to the method through an \texttt{order} edge.}
  \label{fig:ctor_link}
\end{figure}

\begin{figure}[h]
  \centering
  \scriptsize{
\begin{lstlisting}[language=java,basicstyle=\ttfamily,numbers=left,frame = single,framexleftmargin=2.1em, xleftmargin=2.0em]
// a constant with a choice of algorithm
private static String DES_CBC = "DES/CBC/PKCS7Padding";

byte[] process(
  byte[] src, byte[] key, byte[] iv) {
  ...
  Cipher cipher = Cipher.getInstance(DES_CBC);
  ... 
}

\end{lstlisting}
    \caption{Simplified example usage of a \texttt{Cipher}. The \texttt{Cipher} is initialized using the value of a field.}
    \label{fig:cipher_field}
  }
\end{figure}

\vspace{0.2cm}\noindent{\bf Constructors and field initializations.}
We observe that constructors and field initializations provide essential information to identify misuses. 
For example, some fields are immutable (e.g. the name of a cryptographic algorithm, such as “DES” and “AES”, as a string constant, and this information is essential to identifying API misuses related to cryptography),
such as the example in Figure \ref{fig:cipher_field}.
Once assigned a value, they are never reassigned and this value is critical in deciding if the use of an API is a misuse.
To include such information, 
we find statements from blocks of code containing the initialization of fields 
(e.g. constructors, field initialization, and initialization blocks),  
linking them to the graph of methods that use these fields. 
As the execution of the initializations and constructor must occur before the execution of other methods, 
they are joined by \texttt{order} edges in the graph, where the initializations are ordered before the method invocations.
An example is shown in Figure \ref{fig:ctor_link}.

\vspace{0.2cm}\noindent{\bf Subtyping and inheritance.}
The self-usage of members of an API often imply different usage constraints~\cite{sven2019investigating},
for example, one source of false positives was related to self-usage, 
in which a class invokes its own API in its own implementation~\cite{sven2019investigating} 
as API usage within its own implementation may deviate from common usage patterns~\cite{sven2019investigating}. 
For example, a class implementing the \texttt{Iterator} interface may internally call its \textit{next()} method without a check of the \textit{hasNext()} method~\cite{amann2018systematic}.
Usage constraints expected of clients, e.g. guarding method calls or checking a return value, may not be necessary for self-usages.
Other studies also found that API clients may extend the API through inheritance~\cite{zhong2017empirical}.
We try to detect this context by including information about subtyping and inheritance in the Extended AUG (EAUG);
if a class implements a given interface, a correct usage of the API may resemble other usages that implements the same interface.
Thus, the EAUG can model self-usages by generalizing it as a usage contextualized by a particular interface.

\begin{figure*}[]
  \centering
  \begin{subfigure}{.5\textwidth}
    \centering
    \includegraphics[width=\columnwidth]{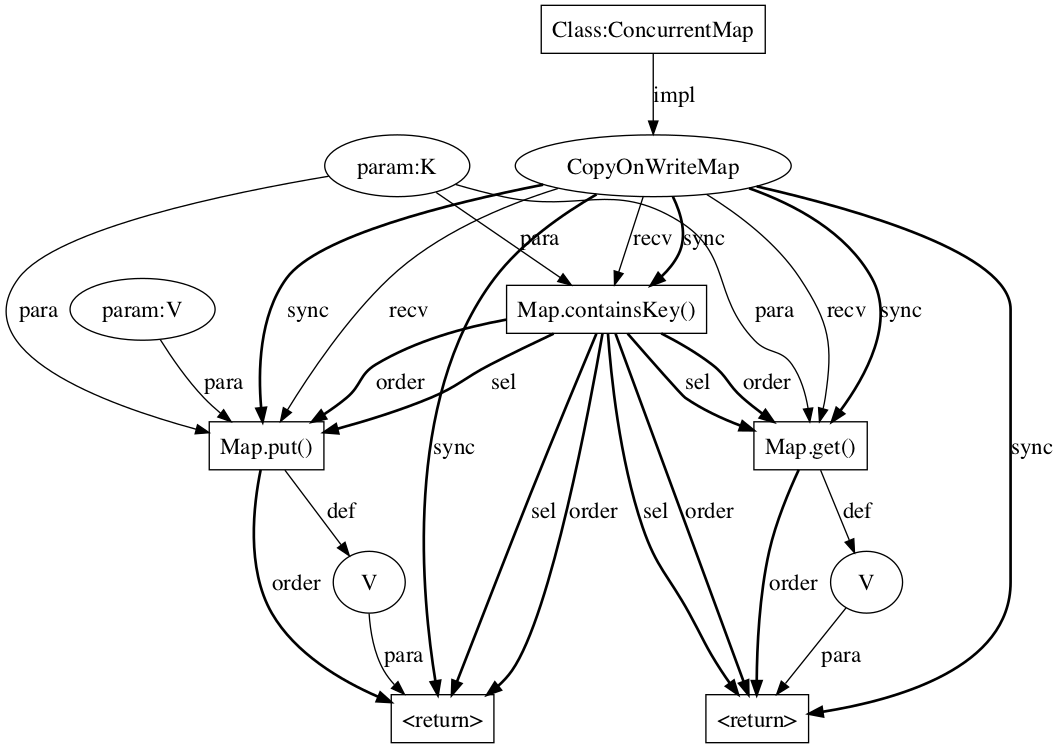}
    \caption{Example of an Extended API Usage Graph}
  \end{subfigure}%
  \begin{subfigure}{.5\textwidth}
    \vspace{0.7cm}
    \centering
    \includegraphics[width=\columnwidth]{Debug-aug-0.png} 
    \caption{Example of an API Usage Graph}
  \end{subfigure}
  \caption{Example of an Extended API Usage Graph, shown in (a). The corresponding API Usage Graph, previously presented in Figure \ref{fig:aug}, of the same program is shown in (b)}
  \label{fig:eaug}
  \end{figure*}

We encode such information by including the superclass/interface 
as a data node linked to the method entry. 
Concretely, this allows \tool{} to extract this information as a subgraph of a single node during the discriminative subgraph mining process.
In the vector representation of the usage example, 
it may not be directly useful on its own, but it may provide useful information in conjunction with the other subgraph features
for the classifier to make a decision.

\vspace{0.2cm}\noindent{\bf Comparison with AUG}
In Figure \ref{fig:eaug}, we provide a visualization of the EAUG of the example shown earlier in Figure \ref{fig:aug_code}, 
which is represented as an AUG shown ealier in Figure \ref{fig:aug}.
In the EAUG, the interface, \texttt{ConcurrentMap}, implemented by the \texttt{CopyOnWriteMap} is captured in a node.
Moreover, the type of the variable is suffixed with ``param:'' if it was passed into the method from the calling site as an argument.
While this example does not use a field, the type of a field would be similarly prefixed with ``field:''.
This allows \tool{} to distinguish between fields, parameters, and locally instantiated variables.

\subsection{Mining GitHub}
\label{sec:github}

To learn subgraph features, we first require a set of usage examples of a given API. 
We take advantage of the numerous usage examples of an API on GitHub.
The examples from GitHub are later labeled by the human annotator and this data is used as the training dataset.

We mine GitHub for usage examples by adapting a tool~\cite{asyrofi2020ausearch} that wraps over GitHub's search API.
Types are resolved on a best-effort basis by downloading the latest version of the libraries whenever requiring third-party dependencies. 
If type resolution fails, we discard the usage example.

As the majority of files on GitHub are clones~\cite{lopes2017dejavu}, we perform a file-level code clone de-duplication
to avoid wasted effort and space storing redundant information.
We use a token-based algorithm similar to SourcererCC~\cite{sajnani2016sourcerercc}. 
Much of SourcererCC focuses on scalability, which we did not need in our work and 
we used only its code comparison algorithm.
Next, methods using the API are identified and a method-level de-duplication is used to further trim the dataset.
Hence, only unique methods will be labeled by a human annotator.

To detect code clones, we use a token-based approach~\cite{sajnani2016sourcerercc}.
We consider a block of code as a bag of tokens. 
The similarity measure, $O\left(B_{x}, B_{y}\right)$, between two code blocks, $O(B_{x}, B_{y})$, is computed based on the number of tokens shared by the code blocks, it is given as follows: \\

\begin{center}
$O\left(B_{x}, B_{y}\right)=\left|B_{x} \cap B_{y}\right|$
\end{center}

If $O\left(B_{x}, B_{y}\right)$ is greater than 0.7, then we consider the code blocks as clones.
As earlier described, we perform code clone deduplication at the method-level, 
and the human annotator does not waste any effort in labeling a code clone of an earlier labeled method.

On average, we find that there are about 7 code clones (explained ) for a file containing an API usage.
37\% of the files have at least one clone, and the most frequently cloned file had over 1600 code clones.
This is in line with previous studies~\cite{lopes2017dejavu,gharehyazie2017some}, which found a large number of copy-and-pasted code on GitHub.

\subsection{Discriminative subgraph mining}
\label{sec:subgraph_mining}

  \label{fig:subgraph_mining}

Unlike typical machine learning applications, graphs cannot be immediately encoded in a vector space, 
which is required to pass the graph as input to a machine learning classifier.
One method of encoding a graph in a vector space is to run a subgraph mining algorithm and identify the best subgraph features.
The graph is represented as a vector indicating if these subgraphs are contained in it.
Then, we mine subgraphs that are \textit{discriminative}, i.e. occurring more frequently in graphs of one label than the other.
Rather than using frequent subgraphs, we posit that discriminative subgraphs are more likely to better represent usage constraints.

\begin{figure}[]

\centering
\begin{tabular}{@{}c@{}}
  \includegraphics[width=\linewidth]{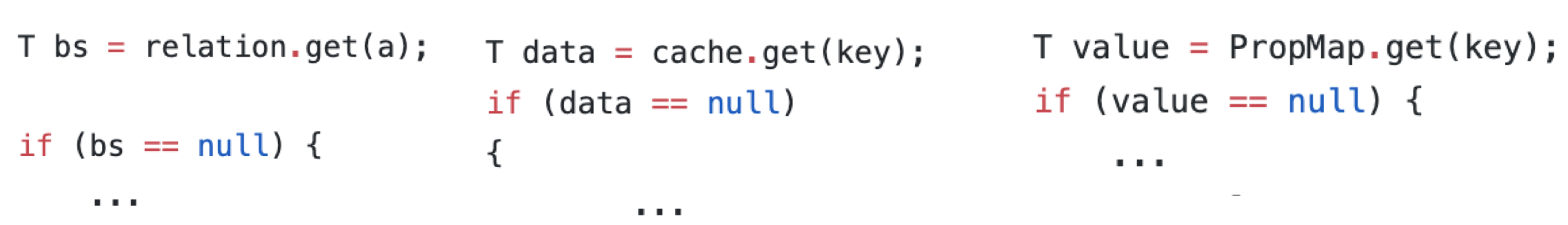} \\[\abovecaptionskip]
  \small (a) Fragments of three examples using \texttt{java.util.Map}
\end{tabular}

\vspace{\floatsep}

\begin{tabular}{@{}c@{}}
  \includegraphics[width=\linewidth]{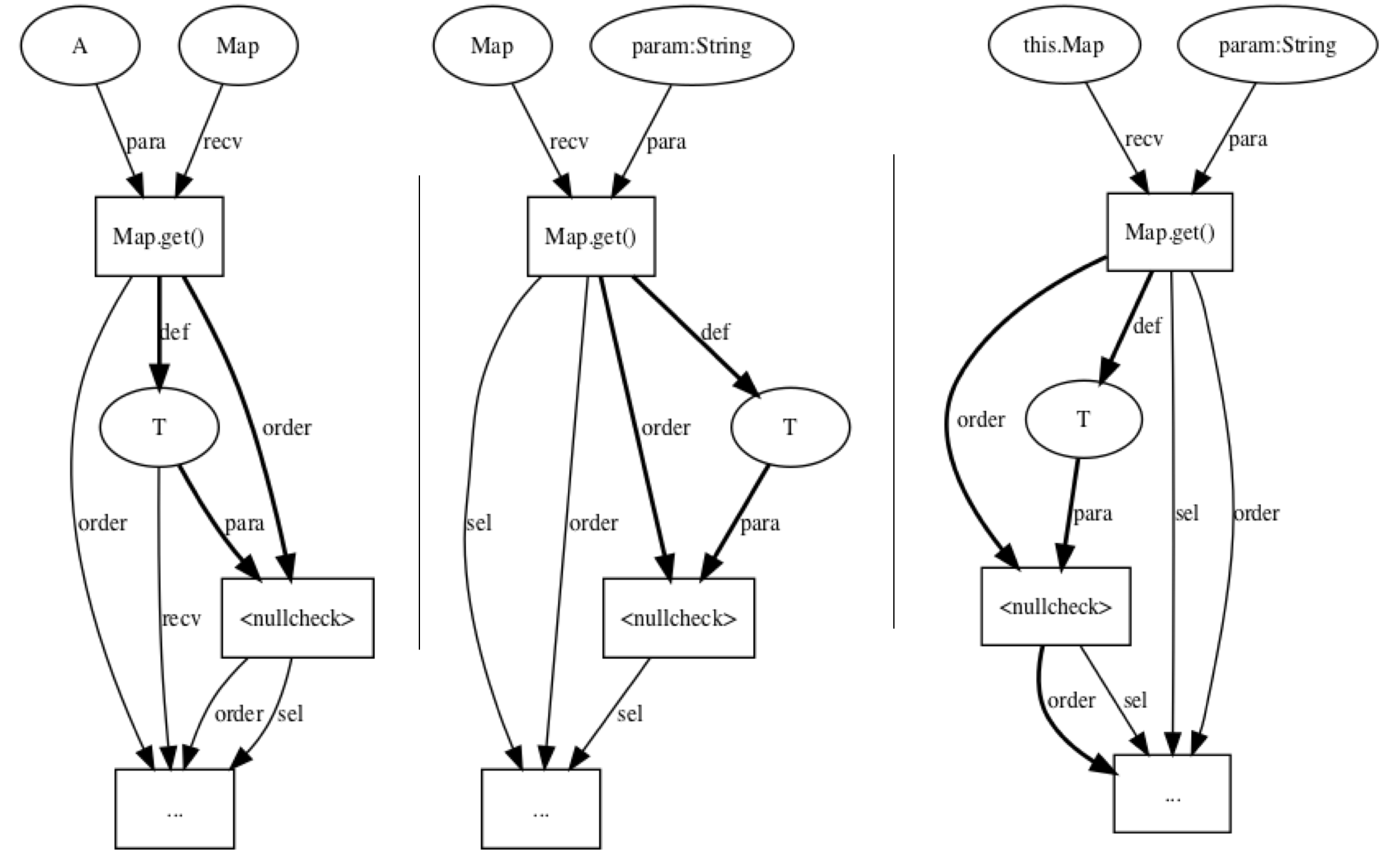} \\[\abovecaptionskip]
  \small (b) Parts of the three examples represented as EAUGs
\end{tabular}
\caption{Simplified examples of three usages of \texttt{java.util.Map}, shown in (a), and parts of their EAUG representation, shown in (b)}
  \label{fig:example_usages}
\end{figure}

\begin{figure}[]

  \includegraphics[width=0.36\columnwidth]{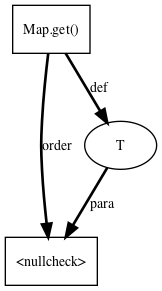}
\centering
\caption{Example discriminative subgraph mined from the examples in Figure \ref{fig:example_usages} after labeling. This discriminative subgraph represents a null-check following \texttt{get}}
  \label{fig:subgraph_map}
\end{figure}

In Figure \ref{fig:example_usages}, we illustrate the discriminative subgraph mining process using the running example.
Given some examples that have been labeled, 
\tool{} mines subgraphs from them.
Among the examples containing a null-check following a \textit{get} method call,
there are significantly more examples that are labeled by the human annotator as correct usages than incorrect usages.
Thus, the subgraph shown in Figure \ref{fig:subgraph_map}, 
common to the three examples in Figure \ref{fig:example_usages}, is identified as a discriminative subgraph feature.

In this study, discriminative subgraphs are identified by two criteria.
As \tool{} enumerates through frequent subgraphs, a test of statistical significance is performed to filter out insignificant subgraphs.
At the end of the frequent subgraph mining process, 
a second round of filtering is done using the CORK criterion~\cite{thoma2009near},
which we use to remove subgraph features that do not contribute to improving classification.

\vspace{0.2cm}\noindent{\bf Enumerating frequent subgraphs}
Enumerating frequent subgraphs allows us to skip the consideration of subgraphs with support below a threshold, $min\_sup$.
However, identifying frequent subgraphs is computationally costly due to the cost of checking for subgraph isomorphism, 
which determines if a graph contains a particular subgraph.
This check is known to be NP-complete.
To enumerate frequent subgraphs quickly, we leverage the frequent subgraph mining algorithm, gSpan~\cite{yan2002gspan}, which is well-understood and efficient.

gSpan takes a collection of graphs and  $min\_sup$ as input, identifying subgraphs with frequency above $min\_sup$ as output. 
To efficiently enumerate the frequent subgraphs, gSpan maps each subgraph to a canonical representation, a minimum DFS code.
Through a depth-first search, gSpan enumerates subgraphs in their DFS code order.
The order of subgraphs visited can be viewed as a traversal of a DFS code tree. 
When a subgraph with a non-minimum DFS code is reached, 
it is directly pruned from the code tree. 
Using this strategy, gSpan visits subgraphs in the canonical search space, 
without the 
computationally expensive test for subgraph isomorphism.
In this work, we set $min\_sup$ to a small value, 3. 
In practice, the threshold required for a subgraph to be discriminative predominantly depends on the Chi-Square test of independence if the choice of the $min\_sup$  parameter is small.

\begin{figure}[ht]

  \includegraphics[width=0.4\columnwidth]{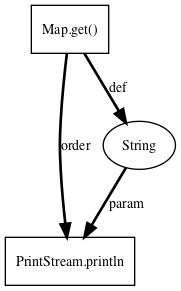}
\centering
\caption{Simplified example of a frequent subgraph that is \textit{not} discriminative, representing usages printing to standard output. \tool{} enumerates such subgraphs, but they fail to pass the checks used to identify discriminative subgraphs. }
  \label{fig:subgraph_map_miss}
\end{figure}

\vspace{0.2cm}\noindent{\bf Testing for significance}
A frequent subgraph may occur commonly among the usage examples of an API, but have no implication 
on the correctness of the usage.
For example, 
a frequent subgraph of \texttt{java.util.Map} may capture the common use of \textit{System.out.println} (as shown in Figure \ref{fig:subgraph_map_miss}),
which is unrelated to using \texttt{java.util.Map},
while another frequent subgraph captures the usage pattern of a null-check following \textit{get}, 
a true usage constraint (previously shown in Figure \ref{fig:subgraph_map}).
Our objective is to filter away the subgraph shown in Figure \ref{fig:subgraph_map_miss}, but keep the subgraph shown in Figure \ref{fig:subgraph_map}.
Therefore, when enumerating the frequent subgraphs, 
we perform a second check that the frequent subgraph has some discriminative power.
In the case of \texttt{java.util.Map}, the subgraph that captures the null-check would be discriminative, 
while the subgraph that encodes printing to \textit{System.out} would be filtered out.

Next, we try to find subgraphs that appear more frequently among one label (C or M).
To determine if a subgraph appears significantly more in graphs of one label than the other, we compute the support of each subgraph.
We count four quantities.
They are $C_H$, the support (number of `hits') of the subgraph among examples labeled C, examples of correct usage, 
$C_M$, the number of graphs labeled C that does not contain the subgraph (`misses’), 
and $M_H$ and $M_M$, similarly defined for the examples labeled M, the examples of misuses.
Then, we use a Chi-Square test of independence to determine if the difference in the number of occurrences of a subgraph in the two set C and M is more often than expected by chance, using a significance level of 0.05 as a threshold.
The Chi-Square statistic measures the statistical significance of a pattern, and is calculated as follows: 
\begin{equation} \label{eq:chisq}
\chi^{2}=\sum_{i=\{C,M\}} \sum_{j=\{H,M\}} \frac{\left(o_{i_{j}}-e_{i_{j}}\right)^{2}}{e_{i_{j}}} 
\end{equation}
$o_{i_{j}}$ is the support observed of $C_H$, $C_M$, $M_H$, and $M_M$.
$e_{i_{j}}$ is the expected support, given the null hypothesis that the subgraph is not discriminative of either label.
Under the null hypothesis, the subgraph is expected to appear a similar number of times in graphs of both labels.
The Chi-Square statistic is, therefore, a measure that the difference in the number of observations did not occur by chance. 
For the null hypothesis to be rejected, the subgraph should appear significantly more frequently in graphs of one label.
The larger $\chi^{2}$ is, the more probable there is a relationship between the subgraph and the label.
Subgraphs that are not discriminative with respect to the two label are filtered out.

\vspace{0.2cm}\noindent{\bf CORK scoring criterion}
Finally, we use the CORK scoring criterion~\cite{thoma2009near}.
While the subgraphs remaining after the previous step are discriminative, 
the subgraphs may not be independent and may frequently co-occur with one another.
The CORK criterion allows us to address this by discarding subgraphs that do not contribute to better classification.
CORK counts the number of \textit{correspondences}, 
which are pairs of misuse and correct examples that cannot be disambiguated from each other given the selected set of features.
A high correspondence indicates that the selected features lack discriminative power and more features should be selected.
CORK is defined by the following equation, where $v$ is the example represented as a feature vector, and $f$ is a feature 
in the set of selected features, $F$. 
Given a pair of labeled examples $i$ and $j$, $i$ and $j$ correspond with each other if they have different labels  but 
their feature vectors, $v$, constructed based on the currently selected features, are identical.

\begin{equation}
  \begin{aligned}
correspondence(i, j) \Leftrightarrow & \left(v^{(i)} \in C\right) \wedge\left(v^{(j)} \in M\right)\\
&  \wedge  \forall f \in F\left(v_{f}^{(i)}=v_{f}^{(j)}\right)
  \end{aligned}
\end{equation}
Initially, $F$ is empty. 
We iterate through the significant subgraphs in the decreasing order of their coverage of the unlabeled dataset, 
considering one subgraph at a time.
A subgraph feature is added to $F$ only if it improves the CORK score (i.e. only if it contributes to disambiguating at least one pair of misuse/correct example).
After this step, we have obtained a set of discriminative subgraphs which will be used to construct the feature vector of a usage site.

\vspace{0.2cm}\noindent{\bf Branch-and-prune}
Mining subgraphs is known to be computationally intensive.
For mining subgraphs in a shorter time, researchers have proposed various methods to speed up the search.
Such approaches may involve pruning the search space. 
For example, gSpan employs a heuristic to prune branches during the traversal of the code tree.
Branches corresponding to supergraphs are pruned  
if a subgraph has frequency below the minimum support, $min\_sup$, 
as any supergraph has a lower frequency than its subgraph.
Similar to existing subgraph mining algorithms~\cite{thoma2009near}, 
\tool{}'s extends the branch-and-prune approach in the canonical search space to speed up subgraph mining.
As our focus is to identify subgraphs that occur more significantly for one label, 
we compute the upper-bound of significance that a supergraph can have in this particular branch. 
For any subgraph, the best case is that one of $C_H$ or $M_H$ is maximized while the other is 0. 
Within the code tree, as we traverse from a subgraph to its supergraph, 
a supergraph feature is most informative when
one of $C_H$ or $M_H$ reaches 0
while the other ($M_H$ or $C_H$) is maintained at its current value. 
As such, we compute the best p-value (as given by the Chi-Square test) that a supergraph can achieve on a traversal of a particular branch. 
If this score is insignificant, then we prune this branch.
This allows us to speed up the subgraph mining process.
In this study, we only consider subgraph patterns with size up to 6 edges to keep running time reasonable.
The size of subgraphs considered influences the running time. 
There is an exponential increase in running time for every increase in the number of edges we consider.
Six was, therefore, selected as this was the maximum size of subgraph that we could mine within one hour from a dataset of over 2000 API usage examples.
The size of the subgraph affects the specificity of the features we mine;
if we set the size to a greater value, then it could mine larger patterns.
However, the larger the pattern is, the more likely it is specific to a few usage locations, 
while smaller patterns are more likely to be found in more usage locations.
Empirically, all the mined subgraphs in our experiments had fewer than 6 edges.

\subsection{Selection of examples to label}
\label{sec:selection}

\tool{} involves a human-in-the-loop, using multiple rounds of labeling.
It initially queries for labels of several dozens randomly sampled unique methods.
For the subsequent rounds of labelling, \tool{} applies principles of active learning~\cite{kong2011dual,ertekin2007learning,huang2010active} to the selection of queries for labeling.
Different from typical scenarios that active learning is applied to~\cite{ertekin2007learning},  
the choice of selected examples to label and the identification of discriminative subgraphs features are closely coupled.
For a subgraph to be identified as a discriminative subgraph, 
it has to occur among the labeled usage examples enough times such that it can be statistically more common in either the correct usage or misuse examples.
Therefore, a sufficient number of graphs that contain the subgraph must be labeled. 
Otherwise this subgraph, regardless of how potentially discriminative it can be (given a hypothetical fully-labeled dataset), cannot be selected as a feature.

However, it is prohibitively expensive for us to label enough graphs to have information about every subgraphs.
As described earlier in Section \ref{sec:background}, performing active learning on graphs requires the 
selection of query graphs in each round.
These query graphs are the usage examples that the human annotator will annotate in the coming round.
Existing studies~\cite{kong2011dual,huang2010active} suggest a good selection of query examples should be both informative and representative. 
A query graph should not be similar to graphs already labeled (i.e. informative),
and it should be similar to other graphs that are unlabeled (i.e. representative).

\vspace{0.2cm}\noindent{\bf Informativeness}
Using these principles, we identify heuristics for picking examples to label.
To pick informative examples, we only select queries from the graphs uncovered by the currently selected features. 
Thus, the query graphs will be dissimilar from already-labeled graphs as they will not share features.
Due to the coupling between the choice of examples to label and subgraph features, 
we propose that the informativeness of labeled examples can be viewed through the informativeness of the subgraphs features they contain.
As such, we measure the informativeness of subgraphs using the notion of coverage following the work of Wang et al.~\cite{wang2013mining}, 
but viewed through the lens of graphs and subgraphs:

\begin{itemize}
\item Coverage: the proportion of total examples containing at least one selected subgraph. 
      Having a high coverage indicates that we can characterize the space of the API usage using the selected subgraphs.
      The more the coverage increases when a subgraph is selected, the more informative the subgraph is.
\end{itemize}

In the example of labeling \texttt{java.util.Map}, if \tool{} has already seen enough examples to know that 
a null-check following \textit{get} is a common and correct usage pattern, 
\tool{} focuses the labeler's attention to examples that do not contain a null-check following \textit{get}.

\vspace{0.2cm}\noindent{\bf Representativeness}
To prevent the selection of non-representative examples, i.e. outliers, 
we filter out subgraphs that appear in too few graphs.
These subgraphs can never be discriminative even if all graphs are labeled, as the number of graphs they appear in is too small.
The number of graphs, $\mathit{min\_signif}$, required to be significant is pre-computed based on the Chi-Square test of independence and the threshold of 
statistical significance (p-value=0.05). 
We pick only query graphs containing the remaining subgraphs.
In other words, 
we  only select graphs containing potentially discriminative subgraphs, 
which are subgraphs that are contained in at least $\mathit{min\_signif }$ graphs. 
Consequently, graphs containing such subgraphs share at least one feature with $\mathit{min\_signif - 1}$ other graphs, and are less likely to be outliers.

To pre-compute $\mathit{min\_signif}$, we simply enumerate possible values starting from 1 and pass these values to the Chi-Square test,
given the number of occurrences of correct and incorrect uses among the currently labeled set of examples.
We pick the smallest value that satisfies the threshold of statistical significance (0.05) and set $\mathit{min\_signif}$ to this value.

\begin{figure}[h]
	\centering
\begin{lstlisting}[language=java,numbers=none,basicstyle=\ttfamily,frame = single]
String rule = inputSourceMap.get(ruleName);
...
InputSource is = new InputSource(
  StreamUtil.stringTOInputStream(rule));
\end{lstlisting}
  \caption{An example usage of \texttt{java.util.Map} that is not representative. It uses project-specific code (\texttt{StreamUtil.stringTOInputStream}). \tool{} has little to gain even if this example was labeled.}
  \label{fig:map_outlier}
\end{figure}

In the example of labeling \texttt{java.util.Map}, \tool{} should avoid wasting the labeler's attention on outliers,
such as uncommon usages that are specific to a given client project (e.g. the example shown in Figure \ref{fig:map_outlier}).
Knowing the labels of the outliers would not help \tool{} to learn discriminative subgraph features.

\vspace{0.2cm}\noindent{\bf Constraint solving}
For each iteration, we view the selection problem as an optimization problem where we pick unlabeled examples to optimize quality metrics with respect to some constraints.
The first constraint is that for each potentially-informative subgraph, we only want to pick the minimal number of graphs for the subgraph to be selected as a discriminative feature.
The second constraint is that, in each batch, we pick at most 0.5\% of graphs to label.
Therefore, solving the optimization problem can be seen as the selection of a minimal number of graphs to maximize our knowledge of subgraphs that cover many graphs.

We encode information about the graphs, subgraphs, and subgraph isomorphisms as a logic program. 
We pass the logic program as input to an off-the-shelf logic program solver, Clingo~\cite{gebser2008user}, 
selected for its ease-of-use and its strong performance against other systems~\cite{calimeri2016design}.
The solution to the optimization problem is the next set of examples to label.
These graphs are labeled and passed to the subgraph mining algorithm.
This process is continued until one of two stopping criteria is satisfied: 1. more than 95\% of the training dataset has been covered by the identified subgraph features,
or 2. we have labeled 5\% of the dataset.
As only 0.5\% of graphs are labelled in each batch, it may take up to 10 batches to satisfy the stopping criterion of 5\% of the dataset being labeled.
These values were picked arbitrarily.
We assumed that a limited number of examples were collected and 5\% was chosen as a sweet spot. 
Empirically, we collected an average of 2330 usage examples per API, and on average, only about 1\% of the examples were labeled.

In this logic program, we consider a set of predicates.
First, we define \texttt{graph(i)} for all uncovered graphs in the dataset.
Next, we define \texttt{subgraph(j)} for all \textbf{frequent} subgraphs that were mined as an intermediate product of the discriminative subgraph mining process.
Note that these are frequent subgraphs, and not discriminative subgraphs, 
as we are trying to determine the next set of graphs to label. 
Therefore, these graphs do not yet contain a subgraph that we have found to be discriminative.
After this, \texttt{covers(i, j)} is defined over \texttt{graph(i)} and \texttt{subgraph(j)},
and it holds subgraph if \texttt{j} is present in the graph \texttt{i}.
As earlier described, \texttt{coverage} gives us the total number of graphs that have been covered; containing at least one subgraph that may be discriminative.
The output of executing the logic program on Clingo is the answer set containing the query graphs, 
which are the graphs that are \texttt{selected}.
These are the usage examples that will be labeled in the next iteration.
Logically, we express the two desired properties of informativeness and representativeness by maximizing the coverage of graphs, 
given a set of subgraphs that may be discriminative subgraphs.

\begin{figure*}[h]
	\centering
	\normalsize{
\begin{lstlisting}[numbers=left,escapechar=!,basicstyle=\ttfamily]
% each graph can be selected only once
{ selected(G) : graph(G) } <= 1 :- graph(G).

% a subgraph may_be_discriminative if n graphs are selected and contains the subgraph
may_be_discriminative(SG) :- #count { 
  G : selected(G), covers(G, SG), graph(G) } >= n, 
  subgraph(SG).

% s or less graphs are selected
:- { selected(G)} > s.

% a graph is covered if a subgraph that may_be_discriminative is contained in it
coverage(X) :- X = #count {G: covers(G,SG), graph(G), may_be_discriminative(SG) }.

#maximize {X : coverage(X) }.
\end{lstlisting}
    \caption{Expressing the constraints as a logic program, trying to find the best \texttt{selected} graphs based on the subgraphs that may be discriminative.}
    \label{fig:clingo_program}
	}
\end{figure*}

We show the encoding of these properties in Figure \ref{fig:clingo_program}.
After the logic program is executed, 
the output are the query graphs, which are the graphs that are \texttt{selected}.
Line 2 specifies a constraint that each graph can be selected for labelling at most once.
Lines 5-7 determine if a subgraph may be discriminative given the currently selected graphs.
A frequent subgraph has to appear a statistically significant number of times (\texttt{n} in Figure \ref{fig:clingo_program}) 
among the labelled graphs for it 
to be considered a discriminative subgraph.
Therefore, if a subgraph appears in fewer labeled graphs than this threshold, it cannot be discriminative.
In this formulation, we consider that a subgraph \textit{may be discriminative} if it appears in at least \texttt{n} query graphs,
where \texttt{n} is the support required for a subgraph to be discriminative.
Line 10 denotes that the total number of query graphs should not exceed \texttt{s}, set to be 0.5\% of the total number of graphs.
This number is the number of instances that the human annotator has to label in the next iteration.
Lines 12 and 15 declares that the choice of \texttt{selected} graphs should maximize the coverage.
In short, we find a small set of graphs such that labelling them may help to identify the set of subgraphs that maximises coverage.

Observe that, at each iteration of selecting a batch of query examples, the coverage can only increase as we only select 
graphs that were not previously covered.
This helps in identifying query graphs that are more informative than if we had randomly selected examples.
Next, the selection of query graph targets only subgraphs that may be discriminative.
This helps to select only graphs that are representative.

\begin{figure}[]

  \includegraphics[width=\columnwidth]{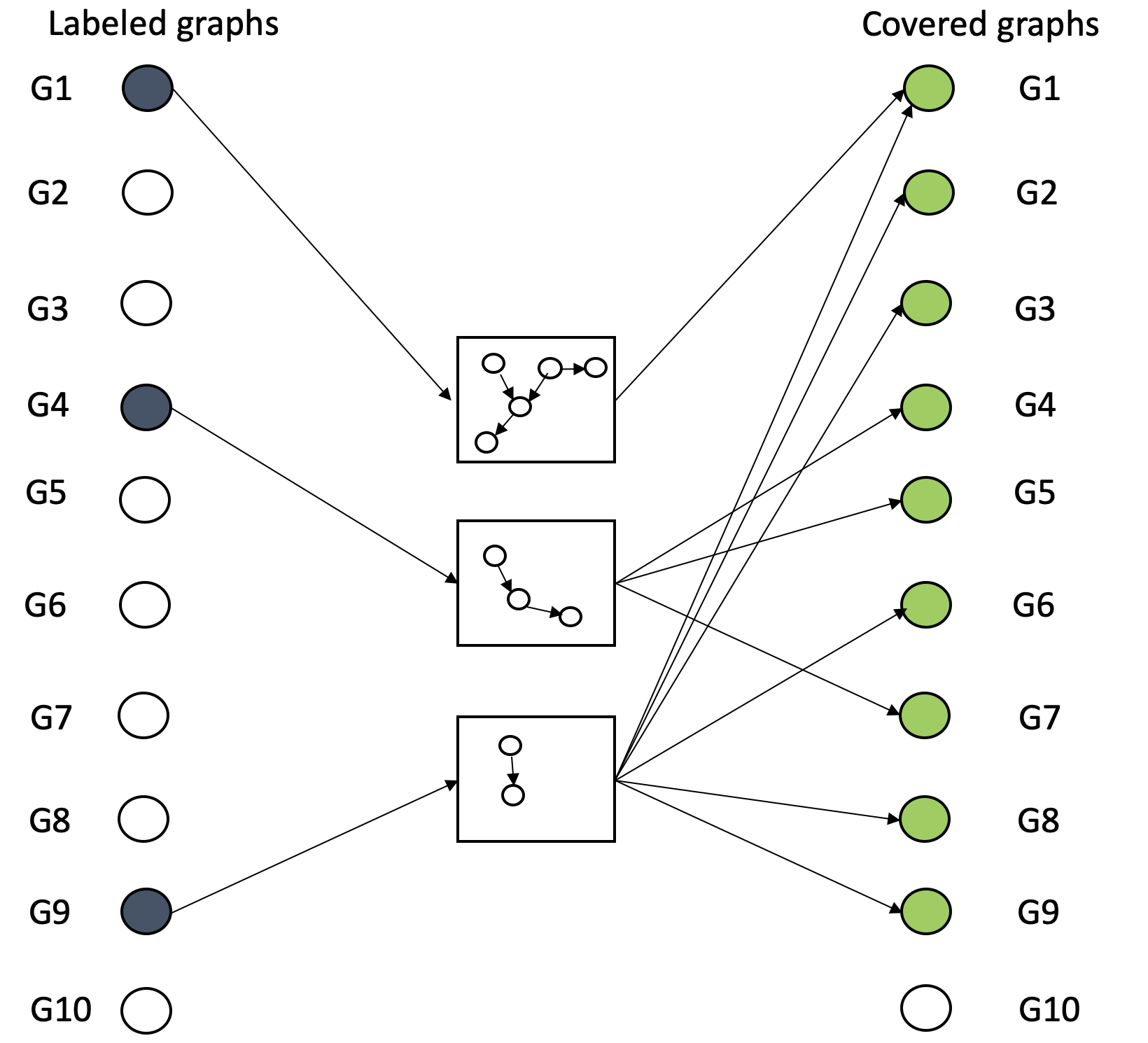}
\centering
\caption{Our objective is to maximise coverage (green dots on the right, each dot represents a one usage example) while minimizing the number of labeled examples (black dots on the left). In this example, by labeling 3 graphs, we gain enough information about the subgraphs that can collectively cover 9 graphs.}
  \label{fig:labeled_to_coverage}
\end{figure}

We visualize this process in Figure \ref{fig:labeled_to_coverage}, 
and to improve the intuition of this process, we work backwards from our objective.
Our objective is to maximise coverage of all unlabeled graphs.
A graph is covered when a subgraph is discriminative and is contained in this graph.
A sub-objective is, then, to determine the choice of discriminative subgraphs that can maximise coverage.
To determine if a graph is discriminative, \tool{} requires a statistically significant number of graphs to be labeled.
In Figure \ref{fig:labeled_to_coverage}, 
to simplify our example,
we assume that only one graph is required to be labeled for a subgraph to be discriminative.
Then, in this simple example, three subgraphs will suffice to cover 9 out of 10 examples. 
The next sub-objective is to select the graph for labeling.
One graph is selected for each potentially discriminative subgraph. 
As there are three subgraphs, three graphs that contain these subgraphs are selected.
By labelling these three selected graphs, 
we will  have information about the subgraphs they contain, and these subgraphs cover most of the examples.

\subsection{Graph classification}
\label{sec:classifying}

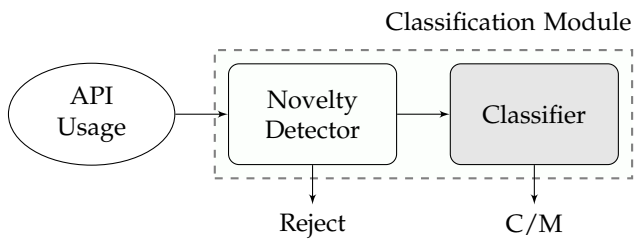
\begin{figure}
  \begin{center}
    
  \begin{tikzpicture}[node distance = 0.7cm , yscale=0.5,
    ,line/.style={draw, -latex'},
    test/.style=%
      {%
        ellipse, draw, text width=4em, text centered, rounded
        corners, minimum height=3em, minimum width=3em
      },
      trace/.style=%
      {%
      rectangle, draw, text width=6em, text centered, rounded
        corners, minimum height=4em, minimum width=4em
      },
      model/.style=%
      {%
      rectangle, draw, fill=gray!20,  text width=6em, text centered, rounded
        corners, minimum height=4em, minimum width=4em
      },
      outer/.style={draw=gray,dashed,fill=green!1,thick,inner sep=5pt },
      drawtext/.style={opacity=0.0,text opacity=1, minimum height=4em},
      miner/.style=%
      {%
      rectangle, draw,   text width=8em, text centered, minimum height=4em, minimum width=4em
      }]
     
      \node[test] (program) {API Usage};

      \node[trace, right = of program] (novelty_detector) {Novelty Detector};

      \node[below = of novelty_detector, yshift=5] (reject) {Reject};

      \node[model, right= of novelty_detector] (classifier) {Classifier};

      \node[below = of classifier, yshift=5] (classification) {C/M};

      \path [line] (program) -- (novelty_detector) ;
      \path [line] (novelty_detector) --  (classifier) ;

      \path [line] (novelty_detector) -- (reject) ;
      \path [line] (classifier) -- (classification) ;

    \begin{pgfonlayer}{background}
      \node[outer,fit=(classifier) (novelty_detector)] (background_0) {};
    \end{pgfonlayer}

    \node[drawtext,text width=4cm, above= of background_0, xshift=1.5cm, yshift=-1cm]  {Classification Module};
  \end{tikzpicture}
    
  \end{center}
  \caption{\tool{}'s Classification Module. Given an input method, it either signals that it cannot be classified (Reject), or classifies it as a misuse (M) or correct use (C).}
  \label{fig:classification}
\end{figure}

The final part of \tool{} is the classification module which uses classification with a reject option.
Apart from (M)isuses and (C)orrect, we introduce a third decision, $reject$, expressing uncertainty and inability to classify the usage instance accurately.
This module comprises of a machine learning classifier and a novelty detector, as seen in Figure \ref{fig:classification}. 
To classify an API usage instance, the input API usage is encoded into an EAUG, $G$, and subgraph isomorphism tests are performed for each discriminative subgraph~\cite{hagberg2008exploring}.
This gives us the feature vector, $v$.

\begin{equation} 
  \label{eq:graph_vector}
  v_{i}=\left\{\begin{array}{ll}
    {1} & {\text { if } F_{i} \sqsubseteq G\left(\text{i.e. }F_{i} \text { is a subgraph of } G\right)} \\
    {0} & {\text { otherwise }}
    \end{array}\right. 
\end{equation}

$v$ is a binary indicator vector, given in Equation \ref{eq:graph_vector}. 
$v_i$ is set to 1 if the graph contains the $i$-th discriminative subgraph feature, $F_{i}$.

To train the classification module, a grid-search is done on the training data to identify the best-performing (in terms of F1) classifier in 5-fold cross validation. 
As the dataset is imbalanced, we oversample instances of the minority class.
SVM (with linear, RBF kernel), K-nearest neighbors 
 and Bayes Classifier (Naive and Complement~\cite{rennie2003tackling}) are included in the grid search.
Our choice of classifiers exclude Deep Learning models as they need voluminous data, 
but our dataset 
typically have only about 50 labels for each API.
We train one model per API.

\vspace{0.2cm}\noindent{\bf Novelty detection}
The novelty detector allows the classification module to identify that a code usage is abnormal and is unrepresented in the training dataset.
Novel usage patterns can appear during practical usage, 
and cannot be characterized by the set of patterns identified from the training dataset.
As we train our classification model, the novelty detector is also tuned on the training dataset. 
To perform novelty detection, we use the Local Outlier Factor~\cite{breunig2000lof} algorithm. 
For each data point, the algorithm measures how isolated is it with respect to its neighbors.
The novelty detector can be viewed as a binary classifier, categorizing an example as an outlier or not.
If it is an outlier, then \tool{} withholds its judgement.
We use an off-the-shelf implementation\footnote{\url{https://scikit-learn.org/stable/modules/generated/sklearn.neighbors.LocalOutlierFactor.html}} of this algorithm in this work.

This algorithm was selected for its effectiveness on data distributions where the data forms clusters of different densities.
We expect that API usage examples follows this distribution.
Due to alternate usage patterns, we expect that majority of examples belong to a core-pattern cluster, 
with smaller, loose clusters containing other examples using alternative patterns. 
An advantage of this technique is that it provides an outlier factor, which is a measure of how much of an outlier a point is, 
rather than just producing a binary output.  
We train one model for each API.

\section{Empirical Evaluation}

\subsection{Benchmarks}

To evaluate the effectiveness of \tool{}, we run an empirical evaluation using two benchmarks.
MUBench was used to evaluate prior work~\cite{amann2016mubench,amann2018systematic,sven2019investigating}, 
and the AU500 dataset that was constructed from API usage instances in real-world projects. 
The training dataset comprises usage examples of the API selected by \tool{} from GitHub,
To evaluate \tool{} on the benchmarks, we took care to construct the training dataset 
such that API usage examples from projects in the benchmarks were not labeled. 
Therefore, \tool{} takes advantage of data from many projects from GitHub, but this dataset is separate from the testing dataset.
For each API, an average of 2330 usage examples (at the granularity of a method) were collected from GitHub.
Using the procedure described in Section \ref{sec:selection}, we use an average of 23 labels for each iteration, 
and an average of slightly over 2 iterations were required for each API. 
This quantity is fewer than other studies that use active learning for tasks related to defect and fault prediction~\cite{thung2015active,lu2012adaptive}, 
as well as studies using active learning for inferring state machine specifications~\cite{walkinshaw2007reverse},
which may require the user to answer several hundred queries.
 
We studied the same APIs as prior work~\cite{sven2019investigating}, 
which are 57 APIs from both the Java standard library as well as third party libraries.
Through the discriminative mining process, we mined about 6 discriminative subgraphs per API.
As the size of the vector passed to the machine learning classifier is equal to the number of discriminative subgraphs,
the average vector size in our experiments is also 6.

All experiments were run on a \textit{Ubuntu} server with an \textit{Intel(R) Core(TM) i7-9700K CPU @ 3.60GHz} and 64GB of RAM.

\subsubsection{MUBench}
First, we perform the experiments using the API Misuse Detector benchmark, MUBench~\cite{amann2016mubench}, 
that was used to systematically evaluate the existing misuse detectors.
These existing misuse detectors are
Tikanga~\cite{wasylkowski2011mining}, Jadet~\cite{wasylkowski2007detecting}, 
DMMC~\cite{monperrus2010detecting}, GrouMiner~\cite{nguyen2009graph}, 
MUDetect~\cite{sven2019investigating},
MUDetectXP (MUDetect on its cross-project setting, where it mines multiple projects~\cite{sven2019investigating}). 
Tikanga, JADET, DMMC, GrouMiner, and MUDetect mine patterns from the project with the misuse.
Instead of only a single project, MUDetectXP uses Boa~\cite{dyer2013boa} to search for examples on GitHub,
mining API usage patterns from multiple projects.
We report the results that were previously reported by Amann et al.~\cite{sven2019investigating}.
Similarly, we use examples from GitHub to train \tool{}. 
While the existing approaches are unsupervised, \tool{} harnesses the power of limited human supervision to boost its performance.

We follow the experimental setup from prior work~\cite{amann2018systematic} for computing Recall and Precision. 
For computing Recall, an experimental procedure known as Experiment R was proposed.
In Experiment R, the findings of the detectors are inspected to count the number of known misuses reported.
The extended version of MUBench was used, containing 208 methods with a misused API. 
Consisting of misuses identified in prior studies, MUBench allows comparison between misuse detectors over these known misuses. 
This procedure allows us to compare and contrast the misuses found by each detector.
We use \tool{} to train models of the APIs studied in Experiment R.

As studies~\cite{flanagan2002extended,kochhar2016practitioners,johnson2013don} have shown that developers rarely use tools producing many false positives,
an experimental setting for computing Precision is provided by MUBench.
In this experimental setup, Experiment P,
ten projects from the MUBench are selected and the misuse detectors are run on these projects. 
We use the same projects as the previous studies: 
Of these ten projects, five were used by Amann et al.~\cite{amann2016mubench} in the original version of MUBench,
while another five were included in the extended version of MUBench~\cite{sven2019investigating}.
These projects were selected as they were among the projects where previously studied detectors could successfully run on, 
and at most one detector did not report any findings~\cite{sven2019investigating}. 

To compute precision, the findings of the misuse detectors are investigated manually. 
Because of the large volume of findings that are reported by existing misuse detectors,  
only the top-20 findings of each detector on each project are investigated,
allowing us to compute Precision@20.
This is the same procedure used for evaluating Precision in previous studies~\cite{amann2018systematic,amann2016mubench,sven2019investigating}.
This keeps the amount of effort required to investigate the findings reasonable.
\tool{} does not produce a ranking of findings.
To allow for comparison against the other detectors, we sample 20 reported misuses using the outlier factors produced by the novelty detector, 
which may be interpreted as a degree of confidence that \tool{} has in a particular finding.
As the outlier factors are not comparable between different APIs, 
we randomly sample (without replacement) 20 misuses inversely weighted by its outlier factor;
we are less likely to sample a misuse if the misuse is likely to be an outlier.

MUBench~\cite{amann2016mubench} also provided an experiment setup for the Recall Upper Bound. 
In this setting, perfect examples to mine patterns of, i.e. only correct usage examples, are provided to the detectors for each misuse.
However, this is irrelevant to \tool{} and MUDetectXP~\cite{sven2019investigating}, 
as they mine GitHub for usage examples.

\subsubsection{AU500}

\begin{table}[t]
  \small
  \caption{Number of usage sites in our evaluation dataset, AU500 }
  \begin{center}
  \begin{tabular}{ |c|l| l | c|}
  \hline
  \textbf{Category} & \textbf{Project} & \textbf{Commit} & \textbf{Sites}
  \\
  \hline
  \multirow{ 4}{*}{GUI} & Apache FOP~\cite{FOP} & 1942336d7& 59
  \\ 
  & SwingX~\cite{swingx} & 820656c &16
  \\ 
  & JFreeChart~\cite{jfreechart}  & 893f9b15 & 51
  \\ 
  & iTextPdf~\cite{itext}  & 2d5b6a212 & 71
  
  \\ 
  \hline
  \multirow{ 4}{*}{Commons} & Apache Lang~\cite{commons-lang}  & 0820c4c89 & 34
  \\ 
  & Apache Math~\cite{commons-math}  & 1abe3c769 & 22
  \\ 
  & Apache Text~\cite{commons-text}  & 7d2b511 & 13
  \\ 
  & Apache BCEL~\cite{commons-bcel} & 5cc4b163 & 3
  \\ \hline
  \multirow{ 4}{*}{Security} & Apache Fortress~\cite{directory-fortress}  & a7ab0c01 & 6
  \\ 
  & Santuario~\cite{santuario-java} & 3832bd83 &15
  \\ 
  & Apache Pdfbox~\cite{pdfbox} & 72249f6ff & 17
  \\ 
  & Wildfly-Eytron~\cite{wildfly}  & a73bbba0f0 &14
  \\ \hline
  \multirow{ 4}{*}{Database} & JackRabbit~\cite{jackrabbit}  & da3fd4199 & 90
  \\ 
  & H2Database~\cite{h2database} & 0ea0365c2 & 84
  \\ 
  & Curator~\cite{curator} &  2af84b9f  & 3
  \\ 
  & Apache BigTop~\cite{bigtop}  & c9cb18fb & 3
  \\ \hline
   & & Total & 500
  \\ \hline
 
  \end{tabular}
  \end{center}
  \label{table:dataset_b_statistic}
\end{table}

To gain a different perspective of the overall effectiveness of \tool{}, 
we construct a new dataset.
To avoid bias from focusing on the misuses collated in MUBench, which were previously identified by existing misuse detectors, 
we sample 500 API usage instances randomly from 16 open-source Java projects, which are clients of the APIs we study.
We use the same client projects as those investigated by Wen et al.~\cite{wen2019exposing} 
These projects were identified
as projects that were from diverse domains, categorized based on their official definition and GitHub topics~\cite{wen2019exposing} . 
Wen et al.~\cite{wen2019exposing} studied the APIs in MUBench, as well as popular APIs discussed on StackOverflow~\cite{zhang2018code}.
However, in this work, we look only for uses of the APIs studied by Amann et al.~\cite{sven2019investigating}, 
as we use MUDetectXP as our baseline.
The details of the projects that are given in Table \ref{table:dataset_b_statistic}.
We use exactly the same projects studied by Wen et al.~\cite{wen2019exposing},
identifying the right commits based on the dates that the projects were accessed by Wen et al.~\cite{wen2019exposing}.
Each usage instance of an API was labeled by human annotators as either a correct use or a misuse.
The dataset and labeling guideline are available on the artifact website.

\begin{table}[t]
  \small
  \caption{Breakdown of correct and misuses in AU500 }
  \begin{center}
  \begin{tabular}{ |l| c | c|}
  \hline
  \textbf{Project} & \textbf{\# Misuses} & \textbf{\# Correct usage} 
  \\
  \hline
  Apache FOP~\cite{FOP} & 18 & 42
  \\ 
  SwingX~\cite{swingx} & 3 & 15
  \\ 
  JFreeChart~\cite{jfreechart}  &  10 & 42
  \\ 
  iTextPdf~\cite{itext}  & 13 & 56
  
  \\ 
  \hline
   Apache Lang~\cite{commons-lang}  & 4 & 30
  \\ 
  Apache Math~\cite{commons-math}  & 3 & 16
  \\ 
  Apache Text~\cite{commons-text}  & 1 & 12
  \\ 
  Apache BCEL~\cite{commons-bcel} & 1 & 2
  \\ \hline
  Apache Fortress~\cite{directory-fortress}  & 0 & 6
  \\ 
  Santuario~\cite{santuario-java} & 2 & 13
  \\ 
  Apache Pdfbox~\cite{pdfbox} & 6 & 11
  \\ 
  Wildfly-Eytron~\cite{wildfly}  & 4 &10
  \\ \hline
  JackRabbit~\cite{jackrabbit}  & 5 & 85
  \\ 
  H2Database~\cite{h2database} & 43 & 41
  \\ 
  Curator~\cite{curator} &  0  & 3
  \\ 
  Apache BigTop~\cite{bigtop}  & 2 & 1
  \\ \hline
  Total & 115 & 385 
  \\ \hline
 
  \end{tabular}
  \end{center}
  \label{table:dataset_b_statistic_breakdown}
\end{table}

As we randomly sampled usages from the clients projects, 
the distribution of API usages in the AU500 reflects the actual distribution of 9972 API usage locations in the client projects.
A further breakdown of the API usages into misuses and correct usages is given in Table \ref{table:dataset_b_statistic_breakdown}.

We had 4 annotators label the dataset. 
Only one of the annotators is an author of this paper.
One of the annotators has a working experience of over 10 years in industry
and all the annotators have at least one year of experience.
Each usage instance was independently labeled by at least 2 annotators.
When there were disagreements, consensus was reached through a discussion between the annotators.

A typical annotator took about 4 minutes to annotate each usage instance, but first spent up to 32 minutes to understand 
the labeling guideline and the APIs through reading its Javadoc, related StackOverflow posts, and examples of misuses.
Similar to previous studies on API misuses, we consider usages at the granularity of a method.
When considering a usage example, an annotator was expected to look at the rest of the source code in the same file.
The usage example is presented in its original source code with code comments intact.
We instructed the annotators to use the projects' Javadoc documentation and code comments of other files in the project when required. 
If a class or method from a third-party library is used, then the annotators were instructed to use the documentation of the third-party library.

For inter-annotator agreement, 
we computed Fleiss' Kappa~\cite{fleiss1971measuring} of 0.49, which is interpreted as Moderate Agreement.
Disagreements were caused by differences in opinion about other mistakes in the same function and if they should be considered 
as a violation of the API's usage constraint. 
Another source of disagreement was over the enclosing class’ undocumented invariants and usage constraints.

\begin{table}[t]
  \centering
  \caption{The number of misuses with a particular violation type, based on the MUC, in the AU500}
  \label{table:muc}
  \begin{tabular}{|l|c|c|c|}
 
 \hline

\textbf{Violation Type}           & \textbf{\# violations} \\ \hline

Missing Method Call  & 51 
\\ \hline
Missing Condition  & 62
\\ \hline
\;\;\;\;null check & 21
\\ \hline
\;\;\;\;value or state  & 41
\\ \hline
Missing Exception Handling  & 2  
\\ \hline
  \end{tabular}
\end{table}

Next, we used the API-Misuse Classification (MUC), developed by Amann et al.~\cite{amann2018systematic} to 
label each misuse with the type of violation. 
The MUC allows for the comparison of different API misuse detectors in terms of the types of violation they can detect.
the MUC distinguishes between misuses with missing or redundant program elements.
The breakdown of violation types in the AU500 is given in Table \ref{table:muc}. 
Compared to MUBench, the AU500 has 4 categories of violation types related to missing API usage elements, 
while MUBench has 7 categories of missing API usage elements.
In the AU500, there are no violations with redundant program elements,
which are the minority of misuses in MUBench.
However, similar to MUBench, the vast majority of misuses in the AU500 are related to missing method calls or condition checks.
While MUBench had a single instance for misuses related to missing or additional synchronization, context, iteration, 
the AU500 does not have any misuses related to them.
In the construction of the AU500, we selected the same APIs as the APIs studied in MUBench.
However, while MUBench largely consists of misuses identified in prior studies,
the instances in the AU500 were randomly sampled from 16 projects.
Therefore, the distribution of misuses in the AU500 are representative of the distribution of misuses in the 16 projects.

Compared to the projects evaluated in MUBench's Experiment P, 
AU500 share 3 projects (IText, JFreeChart, Apache Math), but
different versions (as identified by their commits) of each project were used in the AU500 and MUBench.

For the AU500, we compute Recall, Precision and F-measure by running the misuse detectors on all the projects,
and retain only the reported misuses among the 500 annotated instances.
As every test instance was annotated, 
we can compute Precision instead of computing Precision@20 as we have done for MUBench.
Precision can be computed similarly to its use in machine learning literature, 
in which it is the proportion of true misuses among the reported misuses in the 500 labeled instances.
True positives is the size of the intersection of the reported misuses and the instances labeled as a misuse.
False positives is the size of the intersection of the reported misuses and the instances labeled as a correct usage.
False negatives is the size of the intersection of the reported correct usages and the instances labeled as a misuse.

\begin{center}
  $\text {Precision}=\frac{\text { TP } }{\text {FP}+\text {FP}}$ 
  $\text {Recall}=\frac{\text { TP } }{\text {FP}+\text {FN}}$
\end{center}

Finally, the F1 is the harmonic mean of $Precision$ and $Recall$:
\begin{center}
  $\text {F1}=2 \times \frac{\text { Precision } \times \text {Recall}}{\text {Precision}+\text {Recall}}$
\end{center}

To obtain the results for MUDetectXP, we run MUDetectXP on the 16 projects, but we do not retrain it on new data as it already had models for the APIs we study.
Both \tool{} and MUDetectXP learn their models from data obtained from GitHub.

\subsection{Research Questions}

Our evaluation aims to answer these research questions:

\vspace{0.2cm}\noindent{\bf RQ1. Is \tool{} able to detect misuses previously detected by existing tools?}

This research question concerns the ability of \tool{} to detect misuses found by existing misuse detectors.
We compare \tool{} against existing misuse detectors on MUBench.

\vspace{0.2cm}\noindent{\bf  RQ2. Does \tool{} find more misuses compared to existing approaches?}

The objective of this research question is to evaluate the effectiveness of \tool{}. 
Rather than focusing on the misuses found in previous studies, 
we evaluate \tool{} on the AU500, constructed from API usage instances in real-world projects.

\vspace{0.2cm}\noindent{\bf  RQ3. What is the effect of having a reject option and EAUG in \tool{}? }

In this research question, our objective is to have an ablation study, where we determine if the reject option and our extensions to the AUG are extraneous.
To minimize false positives, \tool{} leverages a novelty detector to reject classifying test usage instances that are dissimilar to training examples.
Also, to incorporate more signals beyond program elements used in the execution of a program, \tool{} used an extended version of the AUG that captures more information.
Using the same metrics, we compare the performance of \tool{} with and without the two components on the AU500 dataset.

\vspace{0.2cm}\noindent{\bf  RQ4. What is the effect of using a different training dataset? }

\tool{} was trained with usage examples of APIs mined from GitHub. 
In machine learning-based approaches, the quality of data used for training is known to affect effectiveness.
Therefore, in our last research question, we investigate the change in performance of \tool{} when a different training dataset is used.
We observe and compare the change in performance metrics when \tool{} is trained on the same dataset used by MUDetectXP~\cite{sven2019investigating}.
This training data was constructed through the 2015 GitHub dataset from BOA~\cite{dyer2013boa}, containing up to 1,000 usage examples for each API.

\subsection{Experimental Results}

\subsubsection{RQ1. Performance of \tool{} on MUBench}

\begin{table}[t]
  \centering
  \caption{Statistics of running MUBench's Experiment P. Other than \tool{}, the results of the detectors were taken from prior work~\cite{sven2019investigating}.}
  \label{table:result_statistic_p}
  \begin{tabular}{|l|c|c|c|}
 
 \hline
\multicolumn{1}{|c|}{\textbf{Detector}}  & \multicolumn{3}{c|}{\textbf{Experiment P}} \\ 
\cline{2-4} 
\multicolumn{1}{|c|}{}                     & 
\textbf{\# findings}           & \textbf{\# true misuses} & \textbf{Prec. @ 20}         \\ \hline

Tikanga  & 85 & 7 & 8.2\%
\\ \hline
JADET  & 91 &8  & 8.8\%
\\ \hline
DMMC & 161 &12 & 7.5\% 
\\ \hline
GrouMiner  &156  & 4 &  2.6\%
\\ \hline
MUDetect  & 146  & 32 & 21.9\%  
\\ \hline
MUDetectXP & 91 & 31 & 34.1\% 
\\ \hline
\textbf{\tool{}} &  164 & 72 & \textbf{43.9}\%  
\\ \hline

  \end{tabular}
  
\end{table}

\begin{table}[t]
  \centering
  \caption{Statistics of running MUBench's Experiment R. \# unique is the number of misuses found only by one detector. Other than \tool{}, the results of the detectors were taken from prior work~\cite{sven2019investigating}.}
  \label{table:result_statistic_r}
  \begin{tabular}{|l|c|c|c|}
 
 \hline
\multicolumn{1}{|c|}{\textbf{Detector}}  & \multicolumn{3}{c|}{\textbf{Misuses in Experiment R}} \\ 
\cline{2-4} 
\multicolumn{1}{|c|}{}                     & 
\textbf{\# found}           & \textbf{\# unique} & \textbf{Recall}    \\ \hline

Tikanga  & 13 & 0 & 6.3\%
\\ \hline
JADET  & 7 & 1 & 3.4\%
\\ \hline
DMMC & 21 & 4 &  10.1\%
\\ \hline
GrouMiner  & 5  & 0 &  2.4\% 
\\ \hline
MUDetect  & 42 & 4 &  20.2\%
\\ \hline
MUDetectXP & 90 & 6 & 43.3\%
\\ \hline

\textbf{\tool{}} & 117 &  35 &  \textbf{56.3}\%
\\ \hline

  \end{tabular}
  
\end{table}

Table \ref{table:result_statistic_p} and \ref{table:result_statistic_r} summarize the results of our evaluation on MUBench. 
During our manual inspection of the results, we find that one of the misuses in MUBench, identified by MUDetect in prior work,
was, in fact, a false positive. 
This case, identified as a misuse of \texttt{java.util.Map}, is actually a usage of Multimap\footnote{\url{https://guava.dev/releases/23.0/api/docs/com/google/common/collect/Multimap.html#get-K-}}. 
Unlike Map, Multimap does not return null, and instead returns an empty collection when \textit{get(key)} is invoked with a key 
not contained in the map. 
As its usage in this case is not a misuse, we removed this case from our evaluation.
Another misuse site was in a project that was no longer publicly accessible, and as such, we also removed this case from our evaluation.
Without using examples from GitHub, MUDetect outperforms the other existing tools in both precision and recall, 
and by using examples from GitHub, MUDetectXP improves substantially over MUDetect and is the strongest baseline.

For Precision@20, following the procedure described in MUBench~\cite{amann2018systematic}, 
we\footnote{Two annotators, including a non-author of this paper who was not informed about the purpose of the labels} manually inspected the top 20 cases found by \tool{} on each project, and finished with a precision of 43.9\%.
We computed Cohen's Kappa to measure the agreement between the two annotators and obtained a Kappa value of 0.75 -- usually interpreted as substantial agreement~\cite{landis1977measurement,sim2005kappa}.  
According to Landis and Koch~\cite{landis1977measurement}, 
a Kappa value between 0.40 to 0.6 corresponds to moderate agreement, while 0.6 to 0.8 is substantial agreement, 
and a value above 0.8 is almost perfect agreement. 
In total, \tool{} reports 164 violations in the top-20 findings in the ten projects, identifying 72 true misuses.
Compared to existing detectors, \tool{} outperforms the baseline detectors in terms of Precision, 
with the strongest baseline, MUDetectXP, achieving a precision of 34.1\%.
Table \ref{table:result_statistic_p} summarizes Experiment P's results. 

For Recall, \tool{} identifies 117 misuses out of the 208 misuse (56.3\%) in MUBench.
In contrast, MUDetectXP identifies 
90 cases out of 208 misuses (43.3\%).
Thus, \tool{} improves over the state-of-the-art by 13\%.

The results of Experiment R are summarized in Table \ref{table:result_statistic_r}.
From the results, there are 35 misuses that only \tool{} can find, 
and there are just 15 misuses that the existing detectors can find that \tool{} is unable to.
This shows that \tool{} is capable of detecting misuses that none of the existing misuse detectors can detect.

Next, we construct a composite baseline detector built from all baselines.
This detector reports a misuse at a particular location as long as a single baseline detector reports a misuse. 
This composite detector would correctly report 111 misuses, and have a resulting Recall of 53.4\%, which is still fewer than the 117 misuses and recall of 56.3\% \tool{} detects.
Comparing the composite detector and \tool{}, there are 15 misuses found only by the composite detector,
while there are 35 misuses found only by \tool{}. 
This indicates that \tool{} is complementary to prior techniques.

Compared to MUDetectXP, \tool{} was able to mine more usage constraints with fewer spurious patterns.
Together with the extensions to the AUG, \tool{} was able to mitigate the effect of self-usages 
(where the internal implementation of an API calls itself) without the need for handcrafted heuristics.
We will provide a further discussion of the differences between \tool{} and MUDetectXP in Section \ref{sec:qualitative}.

We investigated the true misuses that \tool{} did not find.
For misuses related to some APIs, we find that there are too few examples on GitHub to learn any meaningful patterns.
Neither \tool{} nor MUDetectXP was able to detect misuses related to these APIs 
as both correct and incorrect usages of these APIs are too rare.
These APIs include:
\begin{itemize}
\item \textit{org.apache.jackrabbit.core.config.ConfigurationParser}, 
\item \textit{org.apache.commons.lang.text.StrBuilder}, 
\item \textit{org.apache.commons.httpclient.auth.AuthState}.
\end{itemize}

We did not observe any trend regarding the misuses that ALP could detect in the AU500 dataset based on  the categories of misuses in the API-Misuse Classification (MUC)~\cite{amann2018systematic} framework, 
which categorizes misuses based on the type of violation (e.g. missing null-check).
\tool{} was able to detect misuses in the AU500 belonging to every category of the MUC. 
Consistently, the remaining misuses that were undetected by \tool{} are spread across various categories.

\subsubsection{RQ2. Performance of \tool{} on the AU500}
\label{sec:rq2}

Next, we compare the evaluation metrics of the best-performing baseline, 
MUDetectXP~\cite{sven2019investigating}, against \tool{} on the AU500.
As MUDetectXP was shown to outperform the other existing baseline detectors, 
we focus on it.

\begin{table}[t]
  \caption{Experimental results on the AU500 dataset}
  \label{table:dataset_b_statistic_eval}
  \begin{center}
    
  \begin{tabular}{ |c|l|l|l| }
  \hline
   \textbf{Detector} & \textbf{Prec.} & \textbf{Recall } & \textbf{F1}  
  \\
  \hline
  \tool{} & \textbf{44.7}\% & 54.8\% & \textbf{49.2}\% 
  \\ \hline
  \tool{} (w/o reject option) &36.0\% & \textbf{62.6}\% & 45.7\% 
  \\ \hline
  \tool{} (w/o EAUG) & 19.0\% & 45.2\% & 26.7\% 
  \\ \hline
  MUDetectXP  &27.6\% & 29.6\% & 28.6\% 
  \\ \hline
  \end{tabular}
  \end{center}
  
\end{table}

Table \ref{table:dataset_b_statistic_eval} shows the evaluation metrics of \tool{} and MUDetectXP on our evaluation dataset.
Of the 500 usage instances to classify, \tool{} reports 141 of them to be misuses, 
identifying 63 true misuses correctly, 
while MUDetectXP reports misuses for 123 of them, identifying 34 true misuses correctly. 
\tool{} finds 29 more misuses than MUDetect.
In total, there are 115 true misuses among the 500 usage sites.
Therefore, \tool{} has a recall of 54.8\% while MUDetectXP has a recall of 29.6\% (a difference of 25.2\%). 
\tool{} has a precision of 44.7\% while MUDetect has a precision of 27.6\% (a difference of 17.1\%). 
Overall, \tool{} achieves an F1 of 49.2\%, a substantial improvement (of 20.6\%) over MUDetectXP, which achieves an F1 of 28.6\%.

Among the true misuses detected by either tool,
\tool{} managed to find 36 misuses that MUDetectXP did not identify, while MUDetectXP identified 7 misuses that \tool{} did not detect.
There are 27 misuses that both MUDetectXP and ALP identified.
These numbers indicate that if combined, \tool{} and MUDetectXP will detect 70 of the 115 true misuses (a recall of 60.9\%).
This represents an increase in Recall of about 6\% compared to the use of \tool{} alone, and it suggests that
\tool{} is complementary to MUDetectXP.
Overall, \tool{} outperforms the state-of-the-art approach, MUDetectXP, but a developer trying to detect 
as many misuses as possible, regardless of the amount of developer effort needed, should use both tools together.

\subsubsection{RQ3. Effect of reject option and EAUG on \tool{}}

The first two rows of Table \ref{table:dataset_b_statistic_eval} shows the performance of \tool{} with and without the reject option.
Without the reject option, \tool{} reports 59 more instances are buggy (among the 500 annotated usage instances), 
however, its precision is lowered by 8.7\%. 
Among the 59 more findings, only 9 of them were true misuses (or about 15.3\%).
This indicates that the novelty detector was helpful in withholding inaccurate decisions. 
\tool{}'s recall increases by about 7.8\% without the reject option, 
but overall, it attains a reduced F1 of 45.7\% compared to 49.2\%.

On MUBench Experiment R, \tool{} finds 1 more misuse when the reject option is removed.
Without the use of the novelty detector, we have no means to rank the misuses reported by \tool{}.
Therefore, we are unable to compare the change in performance in Precision@20, which requires the top 20 misuses for each project.
Instead, we run \tool{} with and without the reject option, 
and we report the total number of misuses reported among the projects.
Without the reject option, \tool{} reports 543 misuses in total, 
while with the reject option, it reports 440 misuses in total. 
Thus, by removing the reject option, \tool{} would report 23\% more usage locations.
We sampled 30 of the 103 rejected misuses, and found that just 3 of them were true positives. 
Therefore, we see that the reject option helps to mitigate \textbf{Challenge 2}, and overall, improves F1.

Next, we evaluate \tool{} with and without the extensions made to the AUG.
The third row of Table \ref{table:dataset_b_statistic_eval} shows the performance of \tool{} without the EAUG, i.e., using only the original AUG.
We found that the performance of \tool{} dropped substantially without the EAUG,
with a reduction in precision of over 20\% on the AU500.
On MUBench's Experiment R, the Recall of \tool{} dropped to 24.5\%, a decline of 30\%.
We observed that the decrease in effectiveness was caused primarily by reporting many false positives related to \texttt{ResultSet}, 
which implements the \texttt{Closeable} interface.
Without the EAUG distinguishing between parameters and local variables, 
\tool{} would report that any usage of the \texttt{ResultSet} without a corresponding \textit{close} method call as a misuse,
even if the \texttt{ResultSet} was passed in as an argument or was a field.
Therefore, we conclude that capturing this information, as done in the EAUG, 
is important for distinguishing between correct and incorrect usages of some APIs.

\subsubsection{RQ4. Effect of using a different training dataset}
\label{sec:rq4}

\begin{table}[t]
  \caption{Summary of differences in performance metrics when \tool{} is trained with the same dataset (GitHub 2015) as MUDetectXP}
  \vspace{-0.3cm}
  \label{table:dataset_differences}
  \begin{center}
    
  \begin{tabular}{ |c|c|l|l|l|l| }
  \hline

   \multicolumn{1}{|c|}{\textbf{Detector}}  & \multicolumn{2}{c|}{\textbf{MUBench}} & \multicolumn{3}{c|}{\textbf{AU500}} \\ 
\cline{2-6} 
\multicolumn{1}{|c|}{}                     & 

                          \textbf{Prec.@20}  & \textbf{Recall} & \textbf{Prec.}  & \textbf{Recall} & \textbf{F1}   \\ \hline 
  \tool{} (original data) & \textbf{43.9\%} & \textbf{56.3\%} & \textbf{44.7\%} & 54.8\% & \textbf{49.2\%} \\ \hline
  \tool{} (GitHub 2015) & 36.3\% & 47.6\%  & 28.2\% & \textbf{58.3\%}  & 38.0\% \\ \hline
  MUDetectXP  & 34.1\% & 43.3\% & 27.6\% & 29.6\%  & 28.6\% \\ \hline
  
  \end{tabular}
  \end{center}
  
\end{table}

In our final research question, 
we compare the performance of \tool{} when using the same training data as MUDetectXP~\cite{sven2019investigating}.
Referring to the dataset used by MUDetectXP as the GitHub 2015 dataset, 
Table \ref{table:dataset_differences} summarizes the differences in performance metrics.
On both MUBench and the AU500, overall performance declined when the GitHub 2015 dataset was used. 
On MUBench, both Precision@20 and Recall drops from 43.9\% to 36.3\% and from 56.3\% to 47.6\%. 
On the AU500, Precision drops from 44.7\% to 28.2\%. 
While Recall increased from 54.8\% to 58.3\%, the overall performance, in terms of F1, dropped from 49.2\% to 38.0\%.

Compared to MUDetectXP, all performance metrics of \tool{} remained higher. 
In particular, the F1 of \tool{} is almost 10\% higher than the F1 of MUDetectXP on the AU500.
The results suggest that \tool{} improves over MUDetectXP even when using the same training dataset, 
validating our findings and design decisions, even though the choice of training dataset influences overall effectiveness.

Overall, the experimental results suggest that the effectiveness of \tool{} depends 
heavily on having a training dataset that is sufficiently large.
This supports the intuition that \tool{} leverages the diversity in the training dataset to mine discriminative subgraphs and construct an accurate classifier. 
Qualitatively, we observed that the smaller dataset had the consequence of having lower diversity among the usage examples used for training.
This hindered the identification of discriminative subgraphs for many APIs.
For some APIs, there were extremely few examples of incorrect usage in the smaller dataset, preventing \tool{} from learning a model of that API.

\section{Discussion}

\subsection{Qualitative Analysis}
\label{sec:qualitative}
So far, this paper has shown a quantitative evaluation of \tool{}. 
Next, we perform a qualitative evaluation of \tool{}.
In this section, we look into the features that were identified by \tool{} and MUDetectXP. 
We also discuss cases of misuses of the most common APIs, 
which provide insights as to why \tool{} was able to detect some misuses that MUDetectXP missed, 
as well as some cases that \tool{} failed to detect a misuse. 

Developers may implement a class that extends \texttt{Enumeration}.
In their implementation of the \textit{nextElement} method,
the obligation to check for \textit{hasMoreElements} falls on the client of the new class.
Take, for example, the code snippet from a real GitHub project given in Figure \ref{fig:enumerator_alternative_patterns}.
Reporting \textit{new FileInputStream(fileNames.nextElement())} as a misuse of \textit{nextElement} results in a false positive.
\tool{} correctly identifies that the information about implementing the \texttt{Enumeration} interface 
is an important feature for detecting misuses of \textit{nextElement}. 
This is possible as \tool{} includes subtyping in its model of source code, the EAUG.
Therefore, \tool{} is able to correctly detect that this usage is a correct usage, avoiding a false positive.
On the other hand, existing misuse detectors are incapable of encoding any pattern that captures such information, 
showing the difficulty of \textbf{Challenge 3}, which motivates an expressive representation of programs.

\begin{figure}[h]
	\centering
	\scriptsize{
\begin{lstlisting}[language=java,numbers=none,escapechar=!,basicstyle=\ttfamily]
class MyInputStreamEnumerator !\textcolor{red}{implements}!
        !\textcolor{red}{Enumeration$<$FileInputStream$>$}! {
    private Enumeration<String> fileNames;

    public MyInputStreamEnumerator(
            Enumeration<String> fileNames) {
        this.fileNames = fileNames;
    }
    @Override public boolean hasMoreElements() {
        return fileNames.hasMoreElements();
    }
    @Override public FileInputStream nextElement() {
        FileInputStream ret = null;
        try {
            ret = new FileInputStream(
                      fileNames.nextElement());
        } catch(FileNotFoundException ex) { 
            ex.printStackTrace(); 
        }
        return ret;
    }
}
\end{lstlisting}
    \caption{A client of \textit{nextElement}. A salient feature recognised by \tool{} is that the class implements \textit{Enumeration}, highlighted in red, which helps \tool{} to correctly judge that this is not a misuse without the need for handcrafted heuristics.}
    \label{fig:enumerator_alternative_patterns}
	}
\end{figure}

\begin{figure}[b]
  
  \includegraphics[scale=0.5]{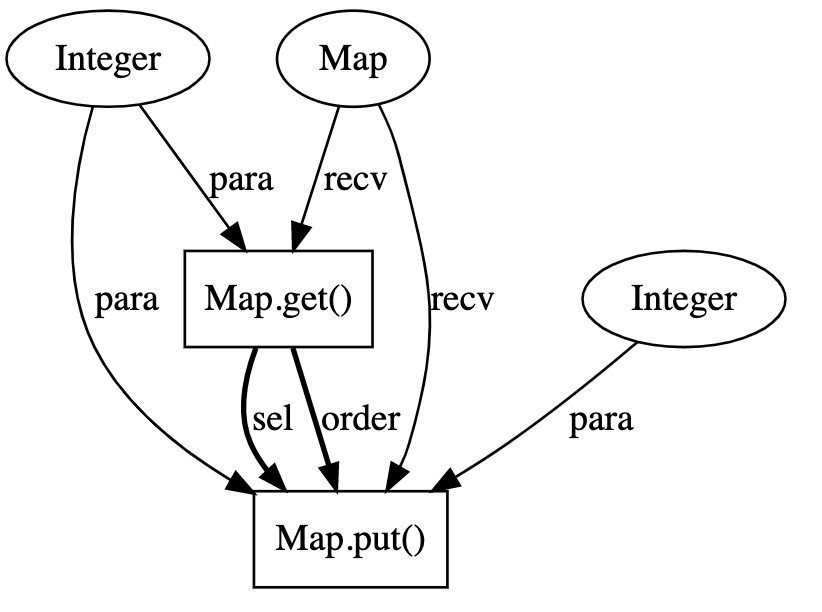}
  \caption{A spurious usage pattern mined by MUDetect. While frequent, its violation leads to no consequences.}
      \label{fig:MD_spurious}
  \centering
  \end{figure}

\begin{table}[t]
  
  \caption{Patterns discovered by MUDetectXP and \tool{}}
  \label{table:pattern_statistics}
  \begin{center}
  \begin{tabular}{ |c|l|l| }
  \hline
    \textbf{Detector} & \textbf{total patterns} & \textbf{\# unrelated}   
  \\
  \hline
  MUDetectXP & 81 & 72
  \\ \hline
  \tool{} & 35 & 9
  \\ \hline
  \end{tabular}
  \end{center}
  
\end{table}

Next, we investigate the patterns mined by \tool{} and MUDetectXP as both MUDetectXP and \tool{} produce subgraph patterns as a by-product.
As each usage instance was compared against the mined patterns, having many patterns that are unrelated to the API usage will lead to lower effectiveness.
We analyze these subgraph patterns for \texttt{java.util.Map}.
As mentioned previously, we focus on \texttt{Map} due to its prevalence in virtually all Java projects.

The results are given in Table \ref{table:pattern_statistics}.
While MUDetectXP did not detect any misuse of \texttt{java.util.Map} in MUBench,
MUDetectXP mines 81 patterns related to it. 
Of these patterns, 72 (88\%) do not relate any of the program elements (e.g. method invocations, parameter values) of Map,
and instead are spurious patterns (e.g. related to printing data to standard output) that occur frequently with Map in the dataset.
These patterns can never identify a misuse of Map, only producing false positives.
The remaining patterns are related to Map, but do not represent usage constraints.
For example, the pattern in Figure \ref{fig:MD_spurious} was identified by MUDetectXP and led to several false positives.
While this pattern frequently appeared, it is not a usage constraint; when violated, there are no consequences.
This demonstrates that MUDetectXP faces \textbf{Challenge 1 and 2}; it is difficult to automatically and reliably determine the correctness of a pattern, 
and when given a violation of a pattern, it is difficult to determine if the violation really is a misuse.

\tool{} mines 35 patterns of \texttt{java.util.Map}.
Of these patterns, just 9 (25\%) of them are patterns that do not show any relationship between program elements of Map.
The patterns mined for each API are available on the artifact website.
\textbf{This suggests that \tool{} addresses Challenge 1 and 2; it reduces the number of spurious patterns that are mined, 
and largely picks up subgraphs that are more likely to affect the correctness of the API usage.}

One reason that \tool{} detects some misuses that are missed by MUDetectXP is that MUDetectXP failed 
to mine the usage constraint using its subgraph mining algorithm.
For example, MUDetectXP is not able to detect misuses related to \texttt{java.util.List}, which was often misused
in cases when its size was not checked before invoking \textit{get(index)} on it.
We hypothesize that this limitation is caused by the different ways that the size can be checked before invoking \textit{get(index)},
for example, within a for-loop or checking for \textit{isEmpty} before invoking \textit{get(0)}, 
as well as the possibility of having arbitrary amounts of code in between the check on the size and the invocation to \textit{get(index)}.
\textbf{On the other hand, \tool{} is able to represent a usage using multiple disjoint subgraphs, 
and the use of human labels allows it to filter out subgraphs that are incidentally mined from the arbitrary code in between the relevant code.}

We also inspected some cases that \tool{} did not successfully detect.
Because \tool{} makes a decision based on  subgraphs
which are disjoint from one another,
it may fail to detect a misuse when a single method contains multiple instances of the object.
For instance, if there are multiple uses of an API in a method,
it cannot detect cases where the first object is correctly used, but the second object is not. 
The first correct usage masks the incorrect second usage.
In contrast, MUDetectXP is able to detect that the second usage violates a usage constraint.

In Section \ref{sec:rq4}, we observed that \tool{} was less effective when using the smaller and, as a result, less diverse dataset.
This is related to another limitation of \tool{}, in which it cannot identify patterns which occur extremely rarely.
Both correct and incorrect usage patterns that are rare cannot be identified as discriminative subgraphs as \tool{} will consider them outliers.
MUDetectXP shares a similar limitation.
If it fails to identify any patterns from a limited set of examples, then it cannot detect any misuses.
However, unlike \tool{}, MUDetectXP only requires examples of correct usages.

Ultimately, while \tool{} outperforms MUDetectXP quantitatively, 
both tools are complementary, outperforming the other under different circumstances.
Fundamentally, both approaches rely on different strategies, having strengths and limitations that are different from each other.
As mentioned previously in Section \ref{sec:rq2} (RQ2),
and they should be used together if the goal is to detect as many misuses as possible.

To conclude, 
our takeaways from the qualitative evaluation are as follows:
\begin{itemize}
  \item \tool{}'s extensions to the API Usage Graph allowed it to correctly detect  self-usages (calling another method from within the same interface) 
  without the need for 
  handcrafted heuristics.
  \item \tool{} has some success in addressing the challenges we described at the start of the paper. It identifies usage constraints while minimizing the number of irrelevant patterns mined.
  \item By representing each graph using a vector of disjoint subgraphs, \tool{} avoids some difficulties faced by prior studies.
  \item \tool{} and MUDetectXP may outperform each other under different circumstances and have different limitations, 
  and to detect as many misuses as possible, 
  both tools should be considered.
\end{itemize}

\subsection{Threats to Validity}

To mitigate threats to internal validity, we double checked our source code and data, however, errors may remain.
Whenever possible, we reused existing, off-the-shelf implementations of algorithms to reduce the risk of implementation mistakes.  
We also make our source code publicly available on the artifact website.
Another threat is bias in our dataset. 
To mitigate this bias, each instance is labeled by at least one annotator with industry experience.

For minimizing threats to construct validity, we reused the same benchmark and evaluation metrics as previous studies. 
For the AU500 dataset, as we have a complete set of labels, we reuse evaluation metrics that are more commonly used for various classification problems in other domains.

The projects used by \tool{} differ from those used by MUDetectXP, 
however, both sets of projects were obtained from GitHub. 
Since the projects were sampled from GitHub, 
we do not expect substantial differences between the projects used. 
However, we note that \tool{} downloaded a larger corpus 
of usage examples (about 2000 examples of API usage examples compared to only 1000 for MUDetectXP). 
When we modified the source code for MUDetectXP to mine patterns from up to 2000 examples, 
we found that MUDetectXP faces scalability issues. 
Running on a machine with 64GB of RAM, MUDetectXP exhausts all the available memory while mining patterns. 
Therefore, we are unable to complete the experiments without changing the subgraph mining process used by MUDetectXP, 
which would fundamentally change how MUDetectXP identifies patterns. 
Furthermore, in RQ4 (see Section \ref{sec:rq4}), we have investigated the performance of \tool{} when using the same dataset as MUDetextXP 
and found that \tool{} still outperformed MUDetectXP on both MUBench and the AU500 dataset. 
Fundamentally, \tool{} mines only discriminative patterns considering informative usage examples, 
while MUDetectXP mines all frequent patterns considering all examples. 

To mitigate threats to external validity,
we have used two benchmarks to increase the spread of API usage examples considered.
A threat is that our findings may not apply to APIs that were not studied. 
However, we study the same APIs as prior work, and both the APIs and projects studied are from diverse domains.

\section{Related Work}

\subsection{API misuses}

\vspace{0.2cm}\noindent{\bf Static analysis}
Many studies have proposed approaches for mining API usage patterns~\cite{zhong2009mapo,li2005pr,wasylkowski2007detecting,wasylkowski2011mining,monperrus2010detecting,nguyen2009graph,sven2019investigating,zhang2018code,thummalapenta2011alattin},
among these studies, several studies describe how to use their patterns for detecting API misuses in source code~\cite{li2005pr,wasylkowski2007detecting,wasylkowski2011mining,monperrus2010detecting,nguyen2009graph,sven2019investigating,thummalapenta2011alattin}. 
Many of these studies mine frequent patterns and detect misuses from anomalies from frequent patterns in their choice of source code representation.
PR-Miner~\cite{li2005pr} uses frequent itemset mining over the function calls invoked to identify association rules between functions.
JADET~\cite{wasylkowski2007detecting} constructs finite state automatas based on temporal properties over method calls and extracts patterns from them. 
It detects anomalies that violate the model and heuristically ranks them.
Building on JADET, Tikanga~\cite{wasylkowski2011mining} converts JADET's models into formulas in Computational Tree Logic,
then use model checking to identify formulae with enough support in a codebase.
DMMC~\cite{monperrus2010detecting} detects missing calls on a receiver type, characterising methods by the invocations on each receiver type.
Alattin~\cite{thummalapenta2011alattin} is an approach that mines alternative patterns to detect missing conditional checks. 
All of these studies proposed unsupervised techniques to mine patterns solely based on their frequency.

Most similar to our work are GrouMiner~\cite{nguyen2009graph} and MUDetect~\cite{sven2019investigating}.
GrouMiner represents programs as groums, which are graphs that encode method calls, field accesses, control, and data-flow.
Frequent pattern mining is used to identify correct usage patterns.
MUDetect~\cite{sven2019investigating} encodes methods as API Usage Graphs, building on top of Groums
but encoding more relationships between program elements, as described earlier in Section \ref{sec:aug}.
MUDetect mines frequent subgraphs through the Apriori algorithm~\cite{inokuchi2000apriori}.
Like our work, MUDetectXP uses examples from GitHub, however, 
unlike \tool{},
it is unsupervised and is limited by the assumption that frequent patterns are correct.
\tool{} is supervised, while identifying a limited number of informative examples to label. 

Compared to other detectors, \tool{} and MUDetectXP mines usage examples from GitHub projects (a cross-project setting) 
rather than a single project.
Both \tool{} and MUDetectXP search for client projects with usage examples using the type of the API, 
and afterwards, use the API to filter usage examples in these client projects.
Without using the type of the API to search and filter for usage examples, 
the pattern mining process will not be scalable as hundreds of projects may be considered for each API.
In contrast, other tools such as GrouMiner~\cite{nguyen2009graph} do not need a target API beforehand.
Instead, they learn usages of the APIs used within a single project and find misuse locations in the same project.

\vspace{0.2cm}\noindent{\bf Dynamic analysis}
Researchers have proposed the use of runtime verification to detect violations of API specifications~\cite{jin2012javamop},
these specifications can be both automatically mined~\cite{pradel2012statically} or written by hand. 
However, Legunsen et al.~\cite{legunsen2016good} have shown that both manually written specifications and automatically mined specifications 
have high false positive rates.
This suggests the need for more tooling in mining and writing API specifications. 
Their analysis revealed that the manually written JavaMOP specifications did not sufficiently encode valid alternative usages. 
\tool{} is complementary to these approaches;
\tool{} can help to discover salient patterns for writing specifications.
Some patterns may not be obvious 
and \tool{} may assist specification writers in discovering unusual usage patterns. 

Catcher~\cite{kechagia2019effective} uses search-based testing to detect API misuses. 
Focusing on the API of the Java standard library, 
it generates test cases directed at code using the API of interest to find a test case triggering an exception from the API.
MUTAPI~\cite{wen2019exposing} is an approach using mutation testing for discovering API misuse patterns.
MUTAPI mutates correct usage of an API within client projects.
In contrast to our work and other API misuse detectors, 
MUTAPI is focused on misuse pattern discovery.
Its precision was not evaluated on MUBench.
While the misuse patterns found by MUTAPI achieved a recall of 49\% on the original version of MUBench~\cite{amann2016mubench} (with about 50 misuses), 
our work achieves a recall of about 56\% on the extended version of MUBench~\cite{amann2018systematic,sven2019investigating} (with 208 misuse locations).
Both Catcher and MUTAPI are limited to misuses which cause exceptions to be thrown,
while our approach can detect misuses with other consequences, such as resource leaks.
We do not directly compare the performance of ALP and MUTAPI 
as MUTAPI relies on the execution of a test suite to detect misuses 
while ALP relies on static analysis to detect misuses. 
MUTAPI can work only on projects with an available test suite, 
while ALP is able to detect misuses in projects without test suites or with test suites with limited coverage. 
Existing research has also shown that many projects have poor code coverage and that developers do not always write test cases\cite{kochhar2015understanding,kochhar2013adoption}.

\vspace{0.2cm}\noindent{\bf Other work on API}

There are other research efforts on other ways to assist development using APIs.
Researchers have worked on better search for API usage examples~\cite{gu2019codekernel,gu2016deep,wang2013mining}, 
recommendation of APIs~\cite{nguyen2016api,nguyen2019focus,fowkes2016parameter}, 
generating or improving documentation~\cite{zhang2019enriching,dekel2009improving}, 
studying API usage in StackOverflow answers~\cite{zhang2018code,reinhardt2018augmenting},
and finding API workarounds~\cite{lamothe2020apis}.
Lamothe et al.~\cite{lamothe2020apis} showed that developers may intentionally use APIs in ways that are not officially supported by the API developers;
these workarounds may be viewed as alternative usage patterns.

\subsection{Active Learning in Software Engineering}

Active learning has been utilized in Software Engineering for various tasks. 
Researchers have used active learning for the generation of assertions from test suite~\cite{pham2017assertion}, 
classification of execution traces~\cite{bowring2004active}, 
and tasks related to defect and fault prediction~\cite{thung2015active,lu2012adaptive}.

Some studies performs active automata learning, using Angluin's L* algorithm~\cite{angluin1987learning} or the QSM technique~\cite{dupont2008qsm}, 
to build models of software systems~\cite{walkinshaw2007reverse,argyros2016sfadiff,de2015protocol}.
For example, the work by Walkinshaw et al.~\cite{walkinshaw2007reverse} use execution traces as input to construct a specification of a software system. 
An oracle, which can be a human user, is involved in answering membership queries, 
i.e. if a sequence of events (e.g. method calls) should be accepted or rejected by the ground-truth state machine.
Others have used active automata learning for finding differences between programs~\cite{argyros2016sfadiff}
and finding security flaws in TLS implementations~\cite{de2015protocol}.
In the study by Walkinshaw et al.~\cite{walkinshaw2007reverse}, the human user answers several hundred queries.
In our work, we do not infer automata.
We use active learning to identify subgraph features for detecting misuses among API usage instances.
Rather than considering just the sequential ordering of events, our work learn patterns over more complex program features.

\section{Conclusion and Future Work}

We propose \tool{}, which represents programs as graphs and classifies them to detect API misuses. 
\tool{} is a human-in-the-loop technique that identifies examples 
for mining discriminative subgraphs.
Through the principles of active learning, a small but informative number of examples are identified and labeled. 
These examples are used to train a classifier with a reject option, 
which reserves judgement when encountering programs that it is uncertain about.
On both MUBench and our newly constructed AU500 dataset, 
\tool{} substantially outperforms existing approaches and the state-of-the-art tool, MUDetectXP.
However, there is still room to improve on both its precision and recall.
In the future, 
we will explore more ways to further improve \tool{} and expand its evaluation to an even larger benchmark with more data.
In particular, we will investigate the effectiveness of \tool{} on other APIs 
as well as further investigate the effects of reducing the amount of labeled data.

\section*{Acknowledgments}
This research/project is supported by the National Research Foundation, Singapore under its Industry Alignment Fund – Pre-positioning (IAF-PP) Funding Initiative. Any opinions, findings and conclusions or recommendations expressed in this material are those of the author(s) and do not reflect the views of National Research Foundation, Singapore.

\ifCLASSOPTIONcaptionsoff
  \newpage
\fi


  \bibliographystyle{IEEEtran}
\bibliography{IEEEabrv,subgraph-api-misuse}

\begin{thebibliography}{100}
\providecommand{\url}[1]{#1}
\csname url@samestyle\endcsname
\providecommand{\newblock}{\relax}
\providecommand{\bibinfo}[2]{#2}
\providecommand{\BIBentrySTDinterwordspacing}{\spaceskip=0pt\relax}
\providecommand{\BIBentryALTinterwordstretchfactor}{4}
\providecommand{\BIBentryALTinterwordspacing}{\spaceskip=\fontdimen2\font plus
\BIBentryALTinterwordstretchfactor\fontdimen3\font minus
  \fontdimen4\font\relax}
\providecommand{\BIBforeignlanguage}[2]{{%
\expandafter\ifx\csname l@#1\endcsname\relax
\typeout{** WARNING: IEEEtran.bst: No hyphenation pattern has been}%
\typeout{** loaded for the language `#1'. Using the pattern for}%
\typeout{** the default language instead.}%
\else
\language=\csname l@#1\endcsname
\fi
#2}}
\providecommand{\BIBdecl}{\relax}
\BIBdecl

\bibitem{egele2013empirical}
M.~Egele, D.~Brumley, Y.~Fratantonio, and C.~Kruegel, ``An empirical study of
  cryptographic misuse in android applications,'' in \emph{Proceedings of the
  2013 ACM SIGSAC conference on Computer \& communications security}, 2013, pp.
  73--84.

\bibitem{nadi2016jumping}
S.~Nadi, S.~Kr{\"u}ger, M.~Mezini, and E.~Bodden, ``Jumping through hoops: Why
  do {Java} developers struggle with cryptography {API}s?'' in
  \emph{Proceedings of the 38th International Conference on Software
  Engineering (ICSE)}, 2016, pp. 935--946.

\bibitem{amann2016mubench}
S.~Amann, S.~Nadi, H.~A. Nguyen, T.~N. Nguyen, and M.~Mezini, ``{MUBench}: a
  benchmark for {API}-misuse detectors,'' in \emph{Proceedings of the 13th
  International Conference on Mining Software Repositories}, 2016, pp.
  464--467.

\bibitem{robillard2012automated}
M.~P. Robillard, E.~Bodden, D.~Kawrykow, M.~Mezini, and T.~Ratchford,
  ``Automated {API} property inference techniques,'' \emph{IEEE Transactions on
  Software Engineering}, vol.~39, no.~5, pp. 613--637, 2012.

\bibitem{legunsen2016good}
O.~Legunsen, W.~U. Hassan, X.~Xu, G.~Ro{\c{s}}u, and D.~Marinov, ``How good are
  the specs? a study of the bug-finding effectiveness of existing {Java} {API}
  specifications,'' in \emph{2016 31st IEEE/ACM International Conference on
  Automated Software Engineering (ASE)}.\hskip 1em plus 0.5em minus 0.4em\relax
  IEEE, 2016, pp. 602--613.

\bibitem{amann2018systematic}
S.~Amann, H.~A. Nguyen, S.~Nadi, T.~N. Nguyen, and M.~Mezini, ``A systematic
  evaluation of static {API}-misuse detectors,'' \emph{IEEE Transactions on
  Software Engineering}, vol.~45, no.~12, pp. 1170--1188, 2018.

\bibitem{sven2019investigating}
------, ``Investigating next steps in static {API}-misuse detection,'' in
  \emph{2019 IEEE/ACM 16th International Conference on Mining Software
  Repositories (MSR)}.\hskip 1em plus 0.5em minus 0.4em\relax IEEE, 2019, pp.
  265--275.

\bibitem{wen2019exposing}
M.~Wen, Y.~Liu, R.~Wu, X.~Xie, S.-C. Cheung, and Z.~Su, ``Exposing library
  {API} misuses via mutation analysis,'' in \emph{2019 IEEE/ACM 41st
  International Conference on Software Engineering (ICSE)}.\hskip 1em plus
  0.5em minus 0.4em\relax IEEE, 2019, pp. 866--877.

\bibitem{thummalapenta2011alattin}
S.~Thummalapenta and T.~Xie, ``Alattin: mining alternative patterns for defect
  detection,'' \emph{Automated Software Engineering}, vol.~18, no. 3-4, pp.
  293--323, 2011.

\bibitem{kruger2019crysl}
S.~Kr{\"u}ger, J.~Sp{\"a}th, K.~Ali, E.~Bodden, and M.~Mezini, ``{CrySL}: An
  extensible approach to validating the correct usage of cryptographic
  {API}s,'' \emph{IEEE Transactions on Software Engineering}, 2019.

\bibitem{gao2019negative}
J.~Gao, P.~Kong, L.~Li, T.~F. Bissyand{\'e}, and J.~Klein, ``Negative results
  on mining crypto-{API} usage rules in {Android} apps,'' in \emph{2019
  IEEE/ACM 16th International Conference on Mining Software Repositories
  (MSR)}.\hskip 1em plus 0.5em minus 0.4em\relax IEEE, 2019, pp. 388--398.

\bibitem{chow1970optimum}
C.~Chow, ``On optimum recognition error and reject tradeoff,'' \emph{IEEE
  Transactions on information theory}, vol.~16, no.~1, pp. 41--46, 1970.

\bibitem{herbei2006classification}
R.~Herbei and M.~H. Wegkamp, ``Classification with reject option,'' \emph{The
  Canadian Journal of Statistics/La Revue Canadienne de Statistique}, pp.
  709--721, 2006.

\bibitem{nguyen2009graph}
T.~T. Nguyen, H.~A. Nguyen, N.~H. Pham, J.~M. Al-Kofahi, and T.~N. Nguyen,
  ``Graph-based mining of multiple object usage patterns,'' in
  \emph{Proceedings of the 7th joint meeting of the European Software
  Engineering Conference and the ACM SIGSOFT symposium on the Foundations of
  Software Engineering (ESEC/FSE)}, 2009, pp. 383--392.

\bibitem{monperrus2010detecting}
M.~Monperrus, M.~Bruch, and M.~Mezini, ``Detecting missing method calls in
  object-oriented software,'' in \emph{European Conference on Object-Oriented
  Programming}.\hskip 1em plus 0.5em minus 0.4em\relax Springer, 2010, pp.
  2--25.

\bibitem{wasylkowski2011mining}
A.~Wasylkowski and A.~Zeller, ``Mining temporal specifications from object
  usage,'' \emph{Automated Software Engineering}, vol.~18, no. 3-4, pp.
  263--292, 2011.

\bibitem{wasylkowski2007detecting}
A.~Wasylkowski, A.~Zeller, and C.~Lindig, ``Detecting object usage anomalies,''
  in \emph{Proceedings of the the 6th joint meeting of the European software
  engineering conference and the ACM SIGSOFT symposium on The foundations of
  software engineering (ESEC/FSE}, 2007, pp. 35--44.

\bibitem{pradel2012statically}
M.~Pradel, C.~Jaspan, J.~Aldrich, and T.~R. Gross, ``Statically checking {API}
  protocol conformance with mined multi-object specifications,'' in \emph{2012
  34th International Conference on Software Engineering (ICSE)}.\hskip 1em plus
  0.5em minus 0.4em\relax IEEE, 2012, pp. 925--935.

\bibitem{zhong2017empirical}
H.~Zhong and H.~Mei, ``An empirical study on {API} usages,'' \emph{IEEE
  Transactions on Software Engineering}, vol.~45, no.~4, pp. 319--334, 2017.

\bibitem{zhong2020para}
H.~Zhong, N.~Meng, Z.~Li, and L.~Jia, ``An empirical study on {API} parameter
  rules,'' in \emph{2020 IEEE/ACM 42th International Conference on Software
  Engineering (ICSE)}, 2020, pp. 899--911.

\bibitem{lin2016lockpeeker}
Z.~Lin, H.~Zhong, Y.~Chen, and J.~Zhao, ``Lockpeeker: detecting latent locks in
  {Java} {API}s,'' in \emph{Proc. ASE}.\hskip 1em plus 0.5em minus 0.4em\relax
  ACM, 2016, pp. 368--378.

\bibitem{dagan1995committee}
I.~Dagan and S.~P. Engelson, ``Committee-based sampling for training
  probabilistic classifiers,'' in \emph{Machine Learning Proceedings
  1995}.\hskip 1em plus 0.5em minus 0.4em\relax Elsevier, 1995, pp. 150--157.

\bibitem{lewis1994sequential}
D.~D. Lewis and W.~A. Gale, ``A sequential algorithm for training text
  classifiers,'' in \emph{SIGIR’94}.\hskip 1em plus 0.5em minus 0.4em\relax
  Springer, 1994, pp. 3--12.

\bibitem{lewis1994heterogeneous}
D.~D. Lewis and J.~Catlett, ``Heterogeneous uncertainty sampling for supervised
  learning,'' in \emph{Machine learning proceedings 1994}.\hskip 1em plus 0.5em
  minus 0.4em\relax Elsevier, 1994, pp. 148--156.

\bibitem{cohn1996active}
D.~A. Cohn, Z.~Ghahramani, and M.~I. Jordan, ``Active learning with statistical
  models,'' \emph{Journal of artificial intelligence research}, vol.~4, pp.
  129--145, 1996.

\bibitem{fujii1998selective}
A.~Fujii, T.~Tokunaga, K.~Inui, and H.~Tanaka, ``Selective sampling for
  example-based word sense disambiguation,'' \emph{Computational
  linguistics-Association for Computational Linguistics}, vol.~24, no.~4, pp.
  573--597, 1998.

\bibitem{settles2008analysis}
B.~Settles and M.~Craven, ``An analysis of active learning strategies for
  sequence labeling tasks,'' in \emph{Proceedings of the 2008 Conference on
  Empirical Methods in Natural Language Processing}, 2008, pp. 1070--1079.

\bibitem{cortes2016learning}
C.~Cortes, G.~DeSalvo, and M.~Mohri, ``Learning with rejection,'' in
  \emph{International Conference on Algorithmic Learning Theory}.\hskip 1em
  plus 0.5em minus 0.4em\relax Springer, 2016, pp. 67--82.

\bibitem{tax2008growing}
D.~M. Tax and R.~P. Duin, ``Growing a multi-class classifier with a reject
  option,'' \emph{Pattern Recognition Letters}, vol.~29, no.~10, pp.
  1565--1570, 2008.

\bibitem{gardner2006one}
A.~B. Gardner, A.~M. Krieger, G.~Vachtsevanos, and B.~Litt, ``One-class novelty
  detection for seizure analysis from intracranial eeg,'' \emph{Journal of
  Machine Learning Research}, vol.~7, no. Jun, pp. 1025--1044, 2006.

\bibitem{perera2019ocgan}
P.~Perera, R.~Nallapati, and B.~Xiang, ``{OCGAN}: One-class novelty detection
  using gans with constrained latent representations,'' in \emph{Proceedings of
  the IEEE Conference on Computer Vision and Pattern Recognition}, 2019, pp.
  2898--2906.

\bibitem{khan2014one}
S.~S. Khan and M.~G. Madden, ``One-class classification: taxonomy of study and
  review of techniques,'' \emph{The Knowledge Engineering Review}, vol.~29,
  no.~3, pp. 345--374, 2014.

\bibitem{chalin2006practitioners}
P.~Chalin, ``Are practitioners writing contracts?'' in \emph{Rigorous
  Development of Complex Fault-Tolerant Systems}.\hskip 1em plus 0.5em minus
  0.4em\relax Springer, 2006, pp. 100--113.

\bibitem{schiller2014case}
T.~W. Schiller, K.~Donohue, F.~Coward, and M.~D. Ernst, ``Case studies and
  tools for contract specifications,'' in \emph{Proceedings of the 36th
  International Conference on Software Engineering (ICSE)}, 2014, pp. 596--607.

\bibitem{hyrumslaw}
``Hyrum's law,'' \url{https://www.hyrumslaw.com/}, accessed 2 Dec 2020.

\bibitem{lamothe2020apis}
M.~Lamothe and W.~Shang, ``When {APIs} are intentionally bypassed: An
  exploratory study of {API} workarounds,'' in \emph{42nd International
  Conference on Software Engineering (ICSE)}, vol. 2020, 2020.

\bibitem{robbes2011study}
R.~Robbes and M.~Lungu, ``A study of ripple effects in software ecosystems
  ({NIER} track),'' in \emph{33rd International Conference on Software
  Engineering}, 2011, pp. 904--907.

\bibitem{moller2020detecting}
A.~M{\o}ller, B.~B. Nielsen, and M.~T. Torp, ``Detecting locations in
  javascript programs affected by breaking library changes,'' \emph{Proceedings
  of the ACM on Programming Languages}, vol.~4, no. OOPSLA, pp. 1--25, 2020.

\bibitem{foo2018efficient}
D.~Foo, H.~Chua, J.~Yeo, M.~Y. Ang, and A.~Sharma, ``Efficient static checking
  of library updates,'' in \emph{26th ACM Joint Meeting on European Software
  Engineering Conference and Symposium on the Foundations of Software
  Engineering (ESEC/FSE)}, 2018, pp. 791--796.

\bibitem{zhang2012automatic}
C.~Zhang, J.~Yang, Y.~Zhang, J.~Fan, X.~Zhang, J.~Zhao, and P.~Ou, ``Automatic
  parameter recommendation for practical {API} usage,'' in \emph{2012 34th
  International Conference on Software Engineering (ICSE)}.\hskip 1em plus
  0.5em minus 0.4em\relax IEEE, 2012, pp. 826--836.

\bibitem{asyrofi2020ausearch}
M.~H. Asyrofi, F.~Thung, D.~Lo, and L.~Jiang, ``{AUSearch}: Accurate {API}
  usage search in github repositories with type resolution,'' in \emph{IEEE
  International Conference on Software Analysis, Evolution and Reengineering},
  2020.

\bibitem{lopes2017dejavu}
C.~V. Lopes, P.~Maj, P.~Martins, V.~Saini, D.~Yang, J.~Zitny, H.~Sajnani, and
  J.~Vitek, ``D{\'e}j{\`a}vu: a map of code duplicates on github,''
  \emph{Proceedings of the ACM on Programming Languages}, vol.~1, no. OOPSLA,
  pp. 1--28, 2017.

\bibitem{sajnani2016sourcerercc}
H.~Sajnani, V.~Saini, J.~Svajlenko, C.~K. Roy, and C.~V. Lopes,
  ``{SourcererCC}: Scaling code clone detection to big-code,'' in
  \emph{Proceedings of the 38th International Conference on Software
  Engineering (ICSE)}, 2016, pp. 1157--1168.

\bibitem{gharehyazie2017some}
M.~Gharehyazie, B.~Ray, and V.~Filkov, ``Some from here, some from there:
  Cross-project code reuse in github,'' in \emph{2017 IEEE/ACM 14th
  International Conference on Mining Software Repositories (MSR)}.\hskip 1em
  plus 0.5em minus 0.4em\relax IEEE, 2017, pp. 291--301.

\bibitem{thoma2009near}
M.~Thoma, H.~Cheng, A.~Gretton, J.~Han, H.-P. Kriegel, A.~Smola, L.~Song, P.~S.
  Yu, X.~Yan, and K.~Borgwardt, ``Near-optimal supervised feature selection
  among frequent subgraphs,'' in \emph{Proceedings of the 2009 SIAM
  International Conference on Data Mining}.\hskip 1em plus 0.5em minus
  0.4em\relax SIAM, 2009, pp. 1076--1087.

\bibitem{yan2002gspan}
X.~Yan and J.~Han, ``{gSpan}: Graph-based substructure pattern mining,'' in
  \emph{2002 IEEE International Conference on Data Mining, 2002.
  Proceedings.}\hskip 1em plus 0.5em minus 0.4em\relax IEEE, 2002, pp.
  721--724.

\bibitem{kong2011dual}
X.~Kong, W.~Fan, and P.~S. Yu, ``Dual active feature and sample selection for
  graph classification,'' in \emph{Proceedings of the 17th ACM SIGKDD
  international conference on Knowledge discovery and data mining}, 2011, pp.
  654--662.

\bibitem{ertekin2007learning}
S.~Ertekin, J.~Huang, L.~Bottou, and L.~Giles, ``Learning on the border: active
  learning in imbalanced data classification,'' in \emph{Proceedings of the
  sixteenth ACM conference on Conference on information and knowledge
  management}, 2007, pp. 127--136.

\bibitem{huang2010active}
S.-J. Huang, R.~Jin, and Z.-H. Zhou, ``Active learning by querying informative
  and representative examples,'' in \emph{Advances in neural information
  processing systems}, 2010, pp. 892--900.

\bibitem{wang2013mining}
J.~Wang, Y.~Dang, H.~Zhang, K.~Chen, T.~Xie, and D.~Zhang, ``Mining succinct
  and high-coverage {API} usage patterns from source code,'' in \emph{2013 10th
  Working Conference on Mining Software Repositories (MSR)}.\hskip 1em plus
  0.5em minus 0.4em\relax IEEE, 2013, pp. 319--328.

\bibitem{gebser2008user}
M.~Gebser, R.~Kaminski, B.~Kaufmann, M.~Ostrowski, T.~Schaub, and S.~Thiele,
  ``A user’s guide to gringo, clasp, clingo, and iclingo,'' 2008.

\bibitem{calimeri2016design}
F.~Calimeri, M.~Gebser, M.~Maratea, and F.~Ricca, ``Design and results of the
  fifth answer set programming competition,'' \emph{Artificial Intelligence},
  vol. 231, pp. 151--181, 2016.

\bibitem{hagberg2008exploring}
A.~Hagberg, P.~Swart, and D.~S~Chult, ``Exploring network structure, dynamics,
  and function using {NetworkX},'' Los Alamos National Lab.(LANL), Los Alamos,
  NM (United States), Tech. Rep., 2008.

\bibitem{rennie2003tackling}
J.~D. Rennie, L.~Shih, J.~Teevan, and D.~R. Karger, ``Tackling the poor
  assumptions of naive bayes text classifiers,'' in \emph{Proceedings of the
  20th international conference on machine learning (ICML)}, 2003, pp.
  616--623.

\bibitem{breunig2000lof}
M.~M. Breunig, H.-P. Kriegel, R.~T. Ng, and J.~Sander, ``{LOF}: identifying
  density-based local outliers,'' in \emph{Proceedings of the 2000 ACM SIGMOD
  international conference on Management of data}, 2000, pp. 93--104.

\bibitem{thung2015active}
F.~Thung, X.-B.~D. Le, and D.~Lo, ``Active semi-supervised defect
  categorization,'' in \emph{2015 IEEE 23rd International Conference on Program
  Comprehension}.\hskip 1em plus 0.5em minus 0.4em\relax IEEE, 2015, pp.
  60--70.

\bibitem{lu2012adaptive}
H.~Lu and B.~Cukic, ``An adaptive approach with active learning in software
  fault prediction,'' in \emph{Proceedings of the 8th International Conference
  on Predictive Models in Software Engineering}, 2012, pp. 79--88.

\bibitem{walkinshaw2007reverse}
N.~Walkinshaw, K.~Bogdanov, M.~Holcombe, and S.~Salahuddin, ``Reverse
  engineering state machines by interactive grammar inference,'' in \emph{14th
  Working Conference on Reverse Engineering (WCRE 2007)}.\hskip 1em plus 0.5em
  minus 0.4em\relax IEEE, 2007, pp. 209--218.

\bibitem{dyer2013boa}
R.~Dyer, H.~A. Nguyen, H.~Rajan, and T.~N. Nguyen, ``Boa: A language and
  infrastructure for analyzing ultra-large-scale software repositories,'' in
  \emph{2013 35th International Conference on Software Engineering
  (ICSE)}.\hskip 1em plus 0.5em minus 0.4em\relax IEEE, 2013, pp. 422--431.

\bibitem{flanagan2002extended}
C.~Flanagan, K.~R.~M. Leino, M.~Lillibridge, G.~Nelson, J.~B. Saxe, and
  R.~Stata, ``Extended static checking for {Java},'' in \emph{ACM SIGPLAN 2002
  Conference on Programming language design and implementation (PLDI)}, 2002,
  pp. 234--245.

\bibitem{kochhar2016practitioners}
P.~S. Kochhar, X.~Xia, D.~Lo, and S.~Li, ``Practitioners' expectations on
  automated fault localization,'' in \emph{25th International Symposium on
  Software Testing and Analysis}, 2016, pp. 165--176.

\bibitem{johnson2013don}
B.~Johnson, Y.~Song, E.~Murphy-Hill, and R.~Bowdidge, ``Why don't software
  developers use static analysis tools to find bugs?'' in \emph{2013 35th
  International Conference on Software Engineering (ICSE)}.\hskip 1em plus
  0.5em minus 0.4em\relax IEEE, 2013, pp. 672--681.

\bibitem{FOP}
``Apache {FOP},'' \url{https://github.com/apache/fop}.

\bibitem{swingx}
``{SwingX},'' \url{https://github.com/ebourg/swingx}.

\bibitem{jfreechart}
``{JFreeChart},'' \url{https://github.com/jfree/jfreechart}.

\bibitem{itext}
``{ITextPDF},'' \url{https://github.com/itext/itextpdf}.

\bibitem{commons-lang}
``{Apache Commons-Lang},'' \url{https://github.com/apache/commons-lang}.

\bibitem{commons-math}
``{Apache Commons-Math},'' \url{https://github.com/apache/commons-math}.

\bibitem{commons-text}
``{Apache Commons-Text},'' \url{https://github.com/apache/commons-text}.

\bibitem{commons-bcel}
``{BCEL},'' \url{https://github.com/apache/commons-bcel}.

\bibitem{directory-fortress}
``{Apache Fortress},'' \url{https://github.com/apache/directory-fortress}.

\bibitem{santuario-java}
``Santuario,'' \url{https://github.com/apache/santuario-java}.

\bibitem{pdfbox}
``{Apache PDFBox},'' \url{https://github.com/apache/pdfbox}.

\bibitem{wildfly}
``{Wildfly-Eytron},'' \url{https://github.com/wildfly-security/wildfly}.

\bibitem{jackrabbit}
``Jackrabbit,'' \url{https://github.com/apache/jackrabbit}.

\bibitem{h2database}
``H2 database,'' \url{https://github.com/h2database/h2database}.

\bibitem{curator}
``{Apache Curator},'' \url{https://github.com/apache/curator}.

\bibitem{bigtop}
``{Apache Bigtop},'' \url{https://github.com/apache/bigtop}.

\bibitem{zhang2018code}
T.~Zhang, G.~Upadhyaya, A.~Reinhardt, H.~Rajan, and M.~Kim, ``Are code examples
  on an online q\&a forum reliable?: a study of {API} misuse on stack
  overflow,'' in \emph{2018 IEEE/ACM 40th International Conference on Software
  Engineering (ICSE)}.\hskip 1em plus 0.5em minus 0.4em\relax IEEE, 2018, pp.
  886--896.

\bibitem{fleiss1971measuring}
J.~L. Fleiss, ``Measuring nominal scale agreement among many raters.''
  \emph{Psychological bulletin}, vol.~76, no.~5, p. 378, 1971.

\bibitem{landis1977measurement}
J.~R. Landis and G.~G. Koch, ``The measurement of observer agreement for
  categorical data,'' \emph{biometrics}, pp. 159--174, 1977.

\bibitem{sim2005kappa}
J.~Sim and C.~C. Wright, ``The kappa statistic in reliability studies: use,
  interpretation, and sample size requirements,'' \emph{Physical therapy},
  vol.~85, no.~3, pp. 257--268, 2005.

\bibitem{zhong2009mapo}
H.~Zhong, T.~Xie, L.~Zhang, J.~Pei, and H.~Mei, ``{MAPO}: Mining and
  recommending {API} usage patterns,'' in \emph{European Conference on
  Object-Oriented Programming}.\hskip 1em plus 0.5em minus 0.4em\relax
  Springer, 2009, pp. 318--343.

\bibitem{li2005pr}
Z.~Li and Y.~Zhou, ``Pr-miner: automatically extracting implicit programming
  rules and detecting violations in large software code,'' \emph{ACM SIGSOFT
  Software Engineering Notes}, vol.~30, no.~5, pp. 306--315, 2005.

\bibitem{inokuchi2000apriori}
A.~Inokuchi, T.~Washio, and H.~Motoda, ``An apriori-based algorithm for mining
  frequent substructures from graph data,'' in \emph{European conference on
  principles of data mining and knowledge discovery}.\hskip 1em plus 0.5em
  minus 0.4em\relax Springer, 2000, pp. 13--23.

\bibitem{jin2012javamop}
D.~Jin, P.~O. Meredith, C.~Lee, and G.~Ro{\c{s}}u, ``{JavaMOP}: Efficient
  parametric runtime monitoring framework,'' in \emph{2012 34th International
  Conference on Software Engineering (ICSE)}.\hskip 1em plus 0.5em minus
  0.4em\relax IEEE, 2012, pp. 1427--1430.

\bibitem{kechagia2019effective}
M.~Kechagia, X.~Devroey, A.~Panichella, G.~Gousios, and A.~van Deursen,
  ``Effective and efficient {API} misuse detection via exception propagation
  and search-based testing,'' in \emph{Proceedings of the 28th ACM SIGSOFT
  International Symposium on Software Testing and Analysis}, 2019, pp.
  192--203.

\bibitem{kochhar2015understanding}
P.~S. Kochhar, F.~Thung, N.~Nagappan, T.~Zimmermann, and D.~Lo, ``Understanding
  the test automation culture of app developers,'' in \emph{2015 IEEE 8th
  International Conference on Software Testing, Verification and Validation
  (ICST)}.\hskip 1em plus 0.5em minus 0.4em\relax IEEE, 2015, pp. 1--10.

\bibitem{kochhar2013adoption}
P.~S. Kochhar, T.~F. Bissyand{\'e}, D.~Lo, and L.~Jiang, ``Adoption of software
  testing in open source projects--a preliminary study on 50,000 projects,'' in
  \emph{2013 17th european conference on software maintenance and
  reengineering}.\hskip 1em plus 0.5em minus 0.4em\relax IEEE, 2013, pp.
  353--356.

\bibitem{gu2019codekernel}
X.~Gu, H.~Zhang, and S.~Kim, ``Codekernel: A graph kernel based approach to the
  selection of {API} usage examples,'' in \emph{2019 34th IEEE/ACM
  International Conference on Automated Software Engineering (ASE)}.\hskip 1em
  plus 0.5em minus 0.4em\relax IEEE, 2019, pp. 590--601.

\bibitem{gu2016deep}
X.~Gu, H.~Zhang, D.~Zhang, and S.~Kim, ``Deep {API} learning,'' in
  \emph{Proceedings of the 2016 24th ACM SIGSOFT International Symposium on
  Foundations of Software Engineering (FSE)}, 2016, pp. 631--642.

\bibitem{nguyen2016api}
A.~T. Nguyen, M.~Hilton, M.~Codoban, H.~A. Nguyen, L.~Mast, E.~Rademacher,
  T.~N. Nguyen, and D.~Dig, ``{API} code recommendation using statistical
  learning from fine-grained changes,'' in \emph{2016 24th ACM SIGSOFT
  International Symposium on Foundations of Software Engineering (FSE)}, 2016,
  pp. 511--522.

\bibitem{nguyen2019focus}
P.~T. Nguyen, J.~Di~Rocco, D.~Di~Ruscio, L.~Ochoa, T.~Degueule, and
  M.~Di~Penta, ``Focus: A recommender system for mining {API} function calls
  and usage patterns,'' in \emph{2019 IEEE/ACM 41st International Conference on
  Software Engineering (ICSE)}.\hskip 1em plus 0.5em minus 0.4em\relax IEEE,
  2019, pp. 1050--1060.

\bibitem{fowkes2016parameter}
J.~Fowkes and C.~Sutton, ``Parameter-free probabilistic {API} mining across
  github,'' in \emph{2016 24th ACM SIGSOFT International Symposium on
  Foundations of Software Engineering (FSE)}, 2016, pp. 254--265.

\bibitem{zhang2019enriching}
J.~Zhang, H.~Jiang, Z.~Ren, T.~Zhang, and Z.~Huang, ``Enriching {API}
  documentation with code samples and usage scenarios from crowd knowledge,''
  \emph{IEEE Transactions on Software Engineering}, 2019.

\bibitem{dekel2009improving}
U.~Dekel and J.~D. Herbsleb, ``Improving {API} documentation usability with
  knowledge pushing,'' in \emph{2009 IEEE 31st International Conference on
  Software Engineering}.\hskip 1em plus 0.5em minus 0.4em\relax IEEE, 2009, pp.
  320--330.

\bibitem{reinhardt2018augmenting}
A.~Reinhardt, T.~Zhang, M.~Mathur, and M.~Kim, ``Augmenting stack overflow with
  {API} usage patterns mined from {GitHub},'' in \emph{2018 26th ACM Joint
  Meeting on European Software Engineering Conference and Symposium on the
  Foundations of Software Engineering}, 2018, pp. 880--883.

\bibitem{pham2017assertion}
L.~H. Pham, L.~L.~T. Thi, and J.~Sun, ``Assertion generation through active
  learning,'' in \emph{International Conference on Formal Engineering
  Methods}.\hskip 1em plus 0.5em minus 0.4em\relax Springer, 2017, pp.
  174--191.

\bibitem{bowring2004active}
J.~F. Bowring, J.~M. Rehg, and M.~J. Harrold, ``Active learning for automatic
  classification of software behavior,'' \emph{ACM SIGSOFT Software Engineering
  Notes}, vol.~29, no.~4, pp. 195--205, 2004.

\bibitem{angluin1987learning}
D.~Angluin, ``Learning regular sets from queries and counterexamples,''
  \emph{Information and computation}, vol.~75, no.~2, pp. 87--106, 1987.

\bibitem{dupont2008qsm}
P.~Dupont, B.~Lambeau, C.~Damas, and A.~v. Lamsweerde, ``The qsm algorithm and
  its application to software behavior model induction,'' \emph{Applied
  Artificial Intelligence}, vol.~22, no. 1-2, pp. 77--115, 2008.

\bibitem{argyros2016sfadiff}
G.~Argyros, I.~Stais, S.~Jana, A.~D. Keromytis, and A.~Kiayias, ``Sfadiff:
  Automated evasion attacks and fingerprinting using black-box differential
  automata learning,'' in \emph{Proceedings of the 2016 ACM SIGSAC conference
  on computer and communications security}, 2016, pp. 1690--1701.

\bibitem{de2015protocol}
J.~De~Ruiter and E.~Poll, ``Protocol state fuzzing of {TLS} implementations,''
  in \emph{24th {USENIX} Security Symposium ({USENIX} Security 15)}, 2015, pp.
  193--206.

\end{thebibliography}

\begin{IEEEbiography}[{\includegraphics[width=1in,height=1.25in,clip,keepaspectratio]{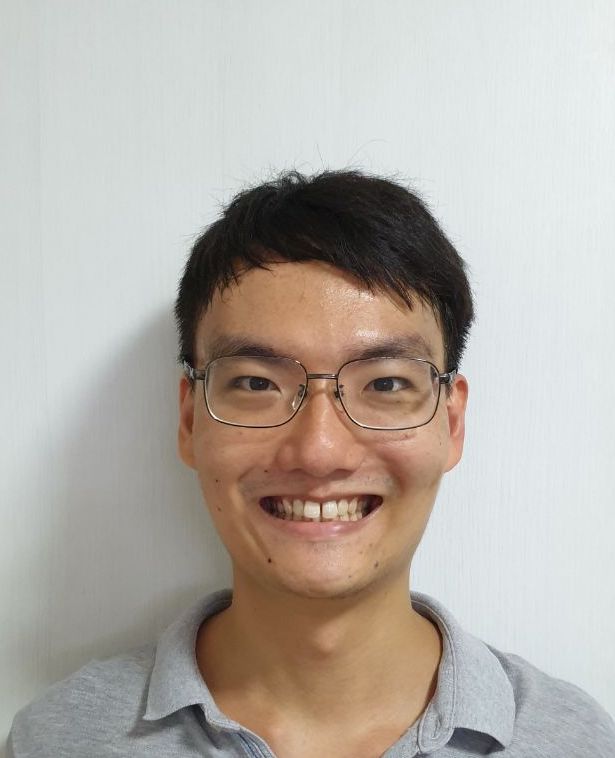}}]{Hong Jin Kang}
is working towards the PhD degree at Singapore Management University. 
He received the bachelor degree in Computer Science from National University of Singapore.
Currently, his research focuses on mining rules and specifications for software engineering.
\end{IEEEbiography}

\begin{IEEEbiography}[{\includegraphics[width=1in,height=1.25in,clip,keepaspectratio]{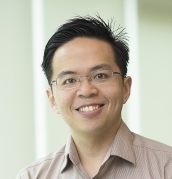}}]{David Lo}
  is an ACM Distinguished Member and a Professor of Computer Science with Singapore Management University, leading the Software Analytics Research (SOAR) group. His research interests include the intersection of software engineering, cybersecurity and data science, encompassing socio-technical aspects and analysis of different kinds of software artefacts, with the goal of improving software quality and security, and developer productivity. His work has been published in major and premier conferences and journals in the area of software engineering, AI, and cybersecurity. By March 2021, he has received more than 15 international research and service awards, including six ACM SIGSOFT Distinguished Paper Awards.
\end{IEEEbiography}

\end{document}